\documentclass[aps,prd,draft,groupedaddress,preprintnumbers,%
              showpacs,showkeys]{revtex4}       
\usepackage[latin1]{inputenc}                    
\usepackage{graphicx}                            
\usepackage{epsfig}
\usepackage{epsf}
\usepackage{latexsym}                            
\usepackage{amsfonts}                            
\usepackage{amssymb}                             
\usepackage{amsmath}                             
\usepackage[mathscr]{eucal}                      
\usepackage{dcolumn}                             
\usepackage{bm}                                  
\usepackage{hyperref}                            

\graphicspath{{.}{Graphics/}}                    

\newcommand{\be}{\begin{equation}}
\newcommand{\ee}{\end{equation}}
\newcommand{\beq}{\begin{eqnarray}}
\newcommand{\eeq}{\end{eqnarray}}
\newcommand{\bea}{\begin{eqnarray}}
\newcommand{\eea}{\end{eqnarray}}

\usepackage{ifpdf}
\usepackage{graphicx}

\def\eq#1{Eq.~(\ref{#1})}

\def \3{\ss }

\newcommand{\tr}{\operatorname{Tr}}
\newcommand{\re}{\operatorname{Re}}

\newcommand{\beqn}{\begin{eqnarray}}
\newcommand{\eeqn}{\end{eqnarray}}

\newcommand{\pslash}{\not\hspace{-3pt}p}
\def\cyp{a}
\def\sac{b}
\def\nic{c}
\def\gre{d}
\def\rmii{e}
\def\mns{f}
\def\ors{h}
\def\liv{g}
\def\ber{j}
\def\zur{i}

\begin{document}

\begin{titlepage}
  {\vspace{-0.5cm} \normalsize
  \hfill \parbox{60mm}{LPT-ORSAY 08-32 \\ 
                       IRFU-08-29\\ 
                        DESY 08-032 \\
                       SFB/CPP-08-19\\
                        ROM2F/2008/06\\
                           HU-EP-08/09\\
                            MS-TP-08-4.\\

}}\\[10mm]
  \begin{center}
    \begin{LARGE}
      \textbf{Light baryon masses with two dynamical twisted
       mass fermions} \\
    \end{LARGE}
  \end{center}

 \vspace{.5cm}

\begin{figure*}[h]
\begin{center}
\epsfxsize=3truecm
\epsfysize=3truecm
    \end{center}
\end{figure*}


 \vspace{-0.8cm}
  \baselineskip 20pt plus 2pt minus 2pt
  \begin{center}
    \textbf{
      C.~Alexandrou$^{(\cyp)}$,
      R.~Baron$^{(\sac)}$, 
      B. Blossier$^{(\nic)}$,
      M.~Brinet$^{(\gre)}$, 
      J.~Carbonell$^{(\gre)}$, 
      P.~Dimopoulos$^{(\rmii)}$,
      V.~Drach$^{(\gre)}$,
      F.~Farchioni$^{(\mns)}$,
      R.~Frezzotti$^{(\rmii)}$,
      P.~Guichon$^{(\sac)}$,
      G.~Herdoiza$^{(\nic)}$,
      K.~Jansen$^{(\nic)}$,
      T.~Korzec$^{(\cyp)}$, 
      G.~Koutsou$^{(\cyp)}$,
      Z.~Liu$^{(\ors)}$, 
      C.~Michael$^{(\liv)}$,
      O.~P\`ene$^{(\ors)}$,
      A.~Shindler$^{(\liv)}$,
      C.~Urbach$^{(\ber)}$,\\
      U.~Wenger$^{(\zur)}$}\\
  \end{center}
  
  \begin{center}
    \begin{footnotesize}
      \noindent 
	
 	$^{(\cyp)}$ Department of Physics, University of Cyprus, P.O. Box 20537,
 	1678 Nicosia, Cyprus
	\vspace{0.2cm}
	
	$^{(\sac)}$CEA-Saclay, IRFU/Service de Physique Nucl\'eaire, 91191 Gif-sur-Yvette, France 
	\vspace{0.2cm}

      $^{(\nic)}$ NIC, DESY, Zeuthen, Platanenallee 6, D-15738 Zeuthen, Germany\\
      \vspace{0.2cm}

	$^{(\gre)}$ Laboratoire de Physique Subatomique et Cosmologie,
               UJF/CNRS/IN2P3, 53 avenue des Martyrs, 38026 Grenoble, France
	 \vspace{0.2cm}

            $^{(\rmii)}$ Dip. di Fisica, Universit{\`a} di Roma Tor Vergata and INFN,
      Sez. di Roma Tor Vergata, Via della Ricerca Scientifica, I-00133 Roma, Italy
      \vspace{0.2cm}

      $^{(\mns)}$ Universit\"at M\"unster, Institut f\"ur Theoretische Physik,
      Wilhelm-Klemm-Strasse 9, D-48149 M\"unster, Germany
      \vspace{0.2cm}

      $^{(\ors)}$ Laboratoire de Physique Th\'eorique (B\^at.~210), Universit\'e
      de Paris XI,CNRS-UMR8627,  Centre d'Orsay, 91405 Orsay-Cedex, France\\
      \vspace{0.2cm}
      
      $^{(\liv)}$ Theoretical Physics Division, Dept. of Mathematical Sciences,
      University of Liverpool, Liverpool L69 7ZL, UK\\
      \vspace{0.2cm}

 $^{(\ber)}$ Humboldt  Universit\"at  zu Berlin, Fachbereich Physik,
             Inst. fur Elementarteilchenphysik, Newtonstr. 15, D-12489 Berlin,
             Germany
      \vspace{0.2cm}
      
      $^{(\zur)}$ Institute for Theoretical Physics, ETH Z{\"u}rich, CH-8093 Z{\"u}rich,
      Switzerland\\
      
    \end{footnotesize}
  \end{center}

  \begin{abstract}
We present results on the mass of the nucleon and  
the $\Delta$ using two
 dynamical degenerate twisted mass quarks and the tree-level Symanzik improved
gauge action. The evaluation
is performed at four quark masses corresponding to a pion mass in the range
of about 300-600~MeV on lattices of 2.1-2.7~fm at three lattice spacings less than 0.1~fm.
We check for cut-off effects by
evaluating these baryon masses on lattices of spatial size 2.1~fm 
at $\beta=3.9$ and $\beta=4.05$
 and on a lattice of 2.4~fm at $\beta=3.8$.
The values we find are compatible  within our statistical errors.
Lattice results are extrapolated to the physical limit using continuum chiral 
perturbation theory.  
Performing a combined fit to our lattice data at $\beta=3.9$ and
$\beta=4.05$ we find a nucleon mass of $963\pm 12 (stat.) \pm 8 (syst.)$~MeV
where we used the lattice spacings determined from the pion decay
constant to convert to physical units.
The
 systematic error  due to the chiral
extrapolation is estimated by comparing results obtained
at  ${\cal O}(p^3)$ 
and  ${\cal O}(p^4)$ heavy baryon chiral perturbation theory.
 The nucleon mass at the physical point provides an independent determination
of the lattice spacing. Using heavy baryon chiral perturbation theory
at ${\cal O}(p^3)$ we find 
$a_{\beta=3.9}=0.0889\pm 0.0012(stat.) \pm 0.0014(syst.)$~fm, 
and $a_{\beta=4.05}= 0.0691\pm 0.0010(stat.) \pm 0.0010(syst.)$~fm, in
good agreement with the values  determined from the pion decay constant. 
Using results from our  two
smaller lattices spacings at constant $r_0 m_\pi$ 
we estimate the continuum limit and check consistency with
results from the  coarser lattice. Results at the continuum limit are 
chirally extrapolated
to the physical point. 
Isospin violating lattice artifacts  in the $\Delta$-system are
found to be compatible with zero  for the values of the  lattice 
spacings used in this work.
Performing a combined fit to our lattice data at $\beta=3.9$ and
$\beta=4.05$ we find  for  the masses of the $\Delta^{++,-}$ and $\Delta^{+,0}$
$1315 \pm 24 (stat.)$~MeV and $1329 \pm 30 (stat.)$~MeV respectively.
 We confirm that in the continuum limit they 
are also degenerate. 
 \end{abstract}
\maketitle 
\end{titlepage}
 
\section{Introduction}
Twisted mass fermions provide an attractive  formulation of lattice QCD that
allows for automatic ${\cal O}(a)$ improvement, infrared regularization 
of small
eigenvalues and fast dynamical 
simulations~\cite{Frezzotti:2000nk,Frezzotti:2004,Jansen:2005,Farchioni:2005,Farchioni:2006}.
A particularly attractive feature is that automatic  
 ${\cal O}(a)$ improvement is obtained by tuning only one parameter
requiring no further improvements on the operator level.
A tree-level analysis of cut-off effects for twisted mass fermions 
has been presented in Ref.~\cite{Cichy:2008}, while a preliminary 
non-perturbative investigation on scaling of several
observables is carried out in Ref.~\cite{Dimopoulos:2007qy}.
Recent simulations
with two
degenerate flavors of
 dynamical Wilson twisted mass fermions demonstrate that  pion masses of 
$m_{\pi}\stackrel{>}{\sim} 300$~MeV can be reached
using Hybrid Monte Carlo methods~\cite{Urbach:2005ji,Jansen:2005yp,Farchioni:2006}. 
The theoretical framework to include the 
strange
and charm quarks has been layed out and  practical simulations are 
being investigated~\cite{Chiarappa:2005mx, Chiarappa:2006ae,Jansen:2007sr}.
Important physical results are emerging using gauge configurations
 generated with two degenerate 
twisted quarks: In the 
 meson sector very precise results on the
pion mass and decay constant led to the determination of the
low energy constants  $\bar{l}_3$, $\bar{l}_4$, F and
$B_0$~\cite{Boucaud:2006,Dimopoulos:2007qy,Urbach:2007} to an accuracy that
had an immediate impact on chiral perturbation theory ($\chi$PT)
 predictions~\cite{Leutwyler:2007}. 
Accurate results on the pion form factor are obtained ~\cite{Simula:2007fa}
 using the ``one-end-trick''
method developed in Refs.~\cite{Foster:1998vw,McNeile:2006bz}.
The kaon system is studied in a partially quenched approach 
by implementing  non-degenerate valence twisted
mass quarks maintaining automatic ${\cal O}(a)$ 
improvement~\cite{Frezzotti:2003xj,Frezzotti:2005ra,AbdelRehim:2006ra,AbdelRehim:2006ve}.
 After determining the
average up and down quark mass and the strange quark 
mass, the kaon decay constant is 
extracted~\cite{Blossier:2007vv,Lubicz:2007iw}. In a similar
approach first results on the charm quark mass and decay constant are
obtained~\cite{Blossier:2007pt}. Preliminary results 
on the first moment of the pion quark distribution 
function were reported in Ref.~\cite{Baron:2007}.

In this work we present a detailed analysis of results in the light baryon 
sector, a subset of which was given in Ref.~\cite{Alexandrou:2007qq}. 
Using two dynamical
 degenerate twisted mass quarks we evaluate the
mass of the nucleon and $\Delta$  for pion  masses down to about 300~MeV.
We use the tree-level Symanzik improved
gauge action~\cite{Weisz:1982}.
We perform the calculation using three different lattice spacings
corresponding to $\beta =3.8$, $\beta=3.9$ and $\beta=4.05$
to check cut-off effects, where $\beta\equiv 6/g^2$ with $g$ being the bare
coupling constant. For each value of $\beta$ we have
configurations
at four different values of the bare quark mass chosen so that
the pion masses are in the range of about 300~MeV  to 600~MeV. 
These gauge configurations belong to the same ensembles as those analyzed for
the evaluation of the pion mass and decay constant. The
values of the lattice spacing extracted  from the pion decay 
constant are 
$a_{\beta=3.8}=0.0995(7)$~fm,  
$a_{\beta=3.9}=0.0855(5)$~fm 
and $a_{\beta=4.05}=0.0667(5)$~fm~\cite{Boucaud:2006, Urbach:2007} 
and will be used in this work.
At  $\beta=3.9$, for the smallest pion mass, 
there are gauge configurations
at two different volumes enabling us to  
 assess  finite volume effects.

Chiral perturbation theory has been successfully applied in the
extrapolation of lattice data obtained with twisted mass fermions
in the pion sector yielding
an accurate determination of the relevant low energy constants.
Applying $\chi$PT to the baryon sector is more involved and several
variants exist.  
However, to leading one-loop order, the result is well established and
the quality of our lattice results allows for extrapolation
to the physical point using this
lowest order result. 
Performing a combined fit to our lattice data at $\beta=3.9$ and
$\beta=4.05$ using the leading one-loop order result 
we find a nucleon mass of $963\pm 12 (stat.)$~MeV,
 where we convert to physical units using the lattice spacing 
determined from $f_\pi$.
We would like to point out that in most other chiral extrapolations
of lattice data the physical point is included in the fits and therefore
such a consistency check cannot be made.
The nucleon mass at the physical point provides
an independent determination of the lattice spacing. We find that
 the lattice spacing thus determined
is in good agreement with the value
extracted in the pion sector. This is a non-trivial check of our lattice
formulation and of the smallness of the systematic errors involved.
To assess systematic errors due to the chiral extrapolation
we perform chiral fits to the nucleon and $\Delta$ mass using higher order
chiral perturbation 
theory
results, which also include explicitly the $\Delta$ degree
of freedom. 

One of the drawbacks of twisted mass fermions is the ${\cal O}(a^2)$ breaking of isospin
symmetry, which is only restored in the continuum limit.  
In the baryon sector we can study isospin breaking 
 by evaluating the mass difference
between $\Delta^{++}(\Delta^-)$ and $\Delta^{+}(\Delta^0)$.
Unlike in the pion sector, where disconnected contributions enter
in the evaluation of the mass of the $\pi^0$, here there are none.
 We can therefore obtain an
accurate evaluation of isospin splitting and its dependence on 
the lattice spacing. We find no isospin splitting within our
statistical accuracy. This is in agreement with a theoretical 
analysis~\cite{Frezzotti:2007,Shindler:2007} that shows
potentially large ${\cal O}(a^2)$ flavor breaking effects to
 appear in the $\pi^0$-mass
but to be suppressed in  other quantities.
Like in the nucleon case, we perform a combined fit to our lattice data at 
$\beta=3.9$ and
$\beta=4.05$ for  the mass of the $\Delta^{++,-}$ and $\Delta^{+,0}$ using the lowest  one-loop order chiral 
perturbation result. We find for the  mass of the $\Delta^{++,-}$ and $\Delta^{+,0}$ 
 $1315 \pm 24 (stat.)$~MeV and $1329 \pm 30 (stat.)$~MeV respectively.
 We confirm that in the continuum limit they 
are also degenerate.

This paper is organized as follows: In Section II we present our lattice
action and in Section III we explain our lattice techniques. In Section IV 
we discuss lattice artifacts
 and in Section V we give results on the nucleon and $\Delta$ mass and also
 describe
the chiral extrapolations. Finally in Section VI we provide a summary and
conclusions.

\section{Lattice action}

For the gauge fields  we use the  tree-level Symanzik improved
gauge action~\cite{Weisz:1982}, which includes besides the
plaquette term $U^{1\times1}_{x,\mu,\nu}$ also rectangular $(1\times2)$ Wilson 
loops $U^{1\times2}_{x,\mu,\nu}$
\begin{equation}
  \label{eq:Sg}
    S_g =  \frac{\beta}{3}\sum_x\Biggl(  b_0\sum_{\substack{
      \mu,\nu=1\\1\leq\mu<\nu}}^4\left \{1-\re\tr(U^{1\times1}_{x,\mu,\nu})\right \}\Bigr. 
     \Bigl.+
    b_1\sum_{\substack{\mu,\nu=1\\\mu\neq\nu}}^4\left \{1
    -\re\tr(U^{1\times2}_{x,\mu,\nu})\right \}\Biggr)\,  
\end{equation}
with  $b_1=-1/12$ and the
(proper) normalization condition $b_0=1-8b_1$. Note that at $b_1=0$ this
action becomes the usual Wilson plaquette gauge action.

The fermionic action for two degenerate flavors of quarks
 in twisted mass QCD is given by
\be
S_F= a^4\sum_x  \bar{\chi}(x)\bigl(D_W[U] + m_0 
+ i \mu \gamma_5\tau^3  \bigr ) \chi(x)
\label{S_tm}
\ee
with   $\tau^3$ the Pauli matrix acting in
the isospin space, $\mu$ the bare twisted mass 
and the massless Wilson-Dirac operator given by 
\be
D_W[U] = \frac{1}{2} \gamma_{\mu}(\nabla_{\mu} + \nabla_{\mu}^{*})
-\frac{ar}{2} \nabla_{\mu}
\nabla^*_{\mu} \quad.
\ee
where
\be
\nabla_\mu \psi(x)= \frac{1}{a}\biggl[U^\dagger_\mu(x)\psi(x+a\hat{\mu})-\psi(x)\biggr]
\hspace*{0.5cm} {\rm and}\hspace*{0.5cm} 
\nabla^*_{\mu}\psi(x)=-\frac{1}{a}\biggl[U_{\mu}(x-a\hat{\mu})\psi(x-a\hat{\mu})-\psi(x)\biggr]
\quad .
\ee
Maximally twisted Wilson quarks are obtained by setting the untwisted quark mass $m_0$ to its critical value $m_{\rm cr}$,
 while the twisted
quark mass parameter $\mu$ is kept non-vanishing in order to work away from the chiral limit.
In \eq{S_tm} the quark fields $\chi$
are in the so-called "twisted basis". The "physical basis" is obtained for
maximal twist by the simple transformation
\be
\psi(x)=\exp\left(\frac {i\pi} 4\gamma_5\tau^3\right) \chi(x),\qquad
\overline\psi(x)=\overline\chi(x) \exp\left(\frac {i\pi} 4\gamma_5\tau^3\right)
\quad.
 \ee
In terms of the physical fields the action is given by
\be
S_F^{\psi}= a^4\sum_x  \bar{\psi}(x)\left(\frac 12 \gamma_\mu 
[\nabla_\mu+\nabla^*_\mu]-i \gamma_5\tau^3 \left(- 
\frac{ar}{2} \;\nabla_\mu\nabla^*_\mu+ m_{\rm cr}\right ) 
+  \mu \right ) \psi(x)\quad.
\label{S_ph}
\ee
In this paper, unless otherwise stated, the quark fields will be understood as ``physical fields'',
 $\psi$, in particular when we define the baryonic interpolating fields. 

A crucial advantage of the twisted mass formulation is
the fact that, by tuning the bare untwisted quark mass $m_0$ to its critical value
 $m_{\rm cr}$, all physical observables are automatically 
${\cal O}(a)$ improved. 
In practice, we implement
maximal twist of Wilson quarks by tuning to zero the bare untwisted current
quark mass, commonly called PCAC mass, $m_{\rm PCAC}$, which is proportional to
$m_0 - m_{\rm cr}$ up to ${\cal O}(a)$ corrections. As detailed in Ref.~\cite{ETMClong}, 
$m_{\rm PCAC}$ is conveniently evaluated through
\be
m_{\rm PCAC}=\lim_{t/a >>1}\frac{\sum_{\bf x}\langle \partial_4 \tilde{A}^b_4({\bf x},t) \tilde{P}^b(0) \rangle}
{2\sum_{\bf x} \langle \tilde{P}^b({\bf x},t)\tilde{P}^b(0)\rangle}, \hspace*{1cm} b=1,2 \quad,
\label{PCAC mass}
\ee
 where $\tilde{A}^b_\mu=\bar{\chi}\gamma_\mu \gamma_5 \frac{\tau^b}{2}\chi$ is the
axial vector current and
 $\tilde{P}^b=\bar{\chi}\gamma_5 \frac{\tau^b}{2}\chi$ 
the pseudoscalar density in the twisted basis. The large $t/a$ limit
is required in order to isolate the contribution of the
lowest-lying charged pseudoscalar meson state in  the correlators of \eq{PCAC mass}. 
This way of determining $m_{\rm PCAC}$ is equivalent
to imposing on the lattice the validity of the axial Ward identity 
$\partial_\mu \tilde{A}_\mu^b = 2m_{\rm PCAC} \tilde{P}^b,\; b=1,2$, 
between the vacuum and the charged zero three-momentum one-pion state.
When $m_0$ is taken such that $m_{\rm PCAC}$ vanishes, this Ward identity
expresses isospin conservation, as it becomes clear by rewriting it in
the physical  quark basis.
The value of $m_{\rm cr}$ is determined at each $\beta$ value at the lowest 
twisted mass, a procedure that preserves ${\cal O}(a)$ improvement
and keeps ${\cal O}(a^2)$ small~\cite{ETMClong,Frezzotti:2005gi}.

The twisted mass fermionic action breaks parity and isospin at 
finite lattice spacing, as it is apparent from the form of the Wilson term in 
 Eq.~(\ref{S_ph}).
In particular,  the isospin breaking in physical observables is a 
cut-off effect of ${\cal O}(a^2)$~\cite{Frezzotti:2004}. However
the up- and down-propagators satisfy
  $G_u(x,y) = \gamma_5 G_d^\dagger(y,x)\gamma_5$, 
which  means that  
two-point correlators are equal with their
hermitian conjugate with u- and d-quarks interchanged.
Using in addition that the  masses are computed from real correlators, it leads 
to the following pairs being
degenerate:
$\pi^+$ and $\pi^-$, proton and
neutron and $\Delta^{++}(\Delta^+)$ and $\Delta^{-}(\Delta^0)$.
 A theoretical analysis~\cite{Frezzotti:2007} shows
that potentially large ${\cal O}(a^2)$ effects that appear in the $\pi^0$-mass
are suppressed in other quantities.
 Calculation of the mass of $\pi^0$, 
which  requires the evaluation of disconnected
diagrams, has been carried out  confirming large ${\cal O}(a^2)$-effects.
In the baryon sector we  study isospin breaking 
 by evaluating the mass difference
between $\Delta^{++}(\Delta^-)$ and $\Delta^{+}(\Delta^0)$.
Since no disconnected contributions enter, 
the mass splitting can be evaluated using fixed source propagators.
An
accurate evaluation of the isospin splitting and its dependence on 
the lattice spacing
is an important component of this work. Examining the size of  isospin
breaking 
is a crucial aspect in particular regarding
future applications of twisted mass fermions to study  baryon structure.
We find that the isospin breaking for the values of the lattice spacing 
considered in this work is consistent with zero within 
our statistical accuracy. Taking the continuum limit of our lattice
results we confirm that
$\Delta^{++,-}$ and $\Delta^{+,0}$ are indeed degenerate leading
to the same mass at the physical point.  

\section{Lattice techniques}
The simulation parameters were chosen such that
the pion mass ranges from about 300-600~MeV. The lattice volumes
and masses used in
this calculation are collected in Table~\ref{Table:params}.
Finite size effects are
examined using the smallest pion mass at $\beta=3.9$ 
as finite volume effects are largest.
At this mass  we have simulations on lattices of 
 spatial size, $L_s\sim 2.1$~fm and $L_s \sim 2.7$~fm.
 
In order to estimate  finite 
  $a$-effects and  the  continuum limit 
  we  use   
two sets of results at $\beta=3.9$  and $\beta=4.05$. Although a further set
of gauge
 configurations at $\beta=3.8$ is analyzed
 this set is not used 
to extrapolate to the continuum limit. 
The reason is that the performance of the HMC algorithm that we use for the simulations
deteriorates when we go to small $\mu$ values on this coarser lattice. 
The long autocorrelation times of the plaquette and the
PCAC mass that we observe~\cite{Boucaud:2006} make the tuning to 
maximal twist less reliable than for the finer lattices.
An analysis aimed at quantifying the impact of possible
 numerical errors from the tuning procedure on observables~\cite{Dimopoulos:2007qy}
is still in progress.
Therefore
the set at $\beta=3.8$ is used 
 only as a cross-check and to estimate  cut-off errors.

\subsection{Interpolating fields}
The masses of the nucleon and the $\Delta$'s are extracted from two-point
correlators using the
standard interpolating fields, which for the proton, 
the $\Delta^{++}$ and $\Delta^{+}$, are given by
\beq \nonumber
J_p &=& \epsilon_{abc}\bigl( u^T_a C\gamma_5 d_b\bigr)u_c, \hspace*{1cm}
J^\mu_{\Delta^{++}} = \epsilon_{abc}\bigl( u^T_a C\gamma^\mu u_b\bigr)u_c, \\
J^\mu_{\Delta^{+}} &=& \frac{1}{\sqrt{3}}\epsilon_{abc}\biggl[
2\bigl( u^T_a C\gamma^\mu d_b\bigr)u_c +\bigl( u^T_a C\gamma^\mu u_b\bigr)d_c
\biggl] ,
\label{interpolate}
\eeq
where $C=\gamma_4\gamma_2$.

\begin{table} 
\begin{tabular}{c|ccccc}
\hline
\multicolumn{6}{c}{ $\beta=4.05$, $a=0.0667(5)$~fm}\\
\hline
$32^3\times 64$, $L_s=2.1$~fm &$a\mu$         & 0.0030     & 0.0060     & 0.0080     & 0.0120\\
                              &$m_\pi$~(GeV) & 0.3070(18) & 0.4236(18) & 0.4884(15) & 0.5981(18) \\
\hline\hline
\multicolumn{6}{c}{$\beta=3.9$, $a=0.0855(5)$~fm}\\
\hline 
 $24^3\times 48$, $L_s=2.1$~fm &$a\mu$         & 0.0040      &   0.0064     &  0.0085     &   0.010 \\ 
                               &$m_\pi$~(GeV) & 0.3131(16)  & 0.3903(9)    & 0.4470(12)  & 0.4839(12)\\
 $32^3\times 64$, $L_s=2.7$~fm &$a\mu$         & 0.0040 & & & \\
                               &$m_\pi$~(GeV) & 0.3082(55) & & & \\
\hline\hline
\multicolumn{6}{c}{ $\beta=3.8$, $a=0.0995(7)$~fm }\\
\hline
$24^3\times 48$, $L_s=2.4$~fm &$a\mu$         & 0.0060     & 0.0080     & 0.0110     & 0.0165\\
                              &$m_\pi$~(GeV) & 0.3667(17) & 0.4128(16) & 0.4799(9) & 0.5855(10) \\
\hline
\end{tabular}
\caption{The parameters of our calculation.}
\label{Table:params}
\end{table}

\begin{figure}[h]
\begin{minipage}{8cm}
\epsfxsize=8truecm
\epsfysize=6truecm
 \mbox{\epsfbox{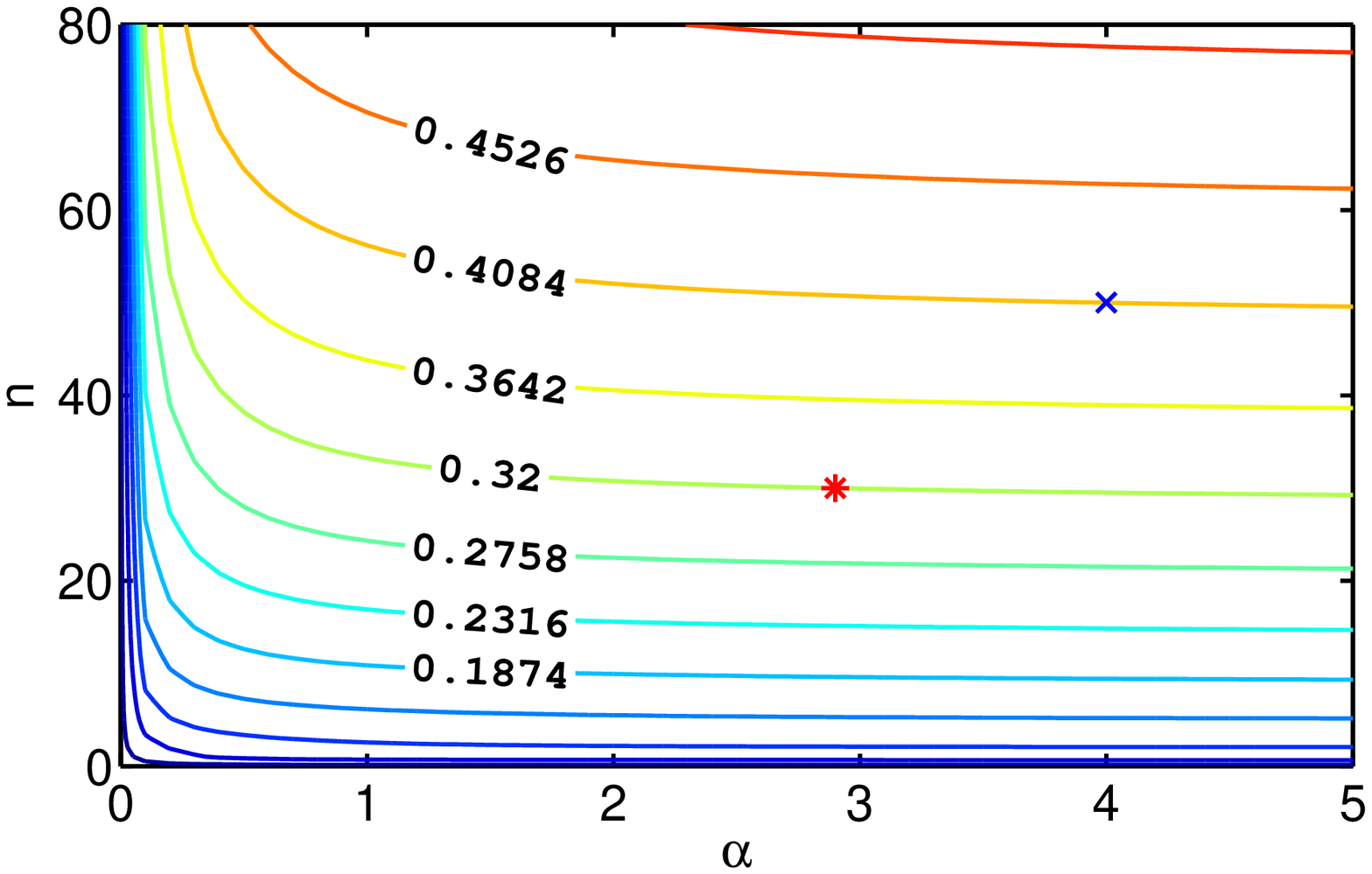}}
\caption{Lines of constant r.m.s radius as a function of the smearing
parameters $\alpha$ and $n$. The asterisk shows the 
values  $\alpha=2.9$, $n=30$ and the
cross  $\alpha=4.0$, $n=50$.\vspace{1.5cm}}
\label{fig:contour}
\end{minipage}
\hfill
\begin{minipage}{8cm}
\epsfxsize=8truecm
\epsfysize=6truecm
 \mbox{\epsfbox{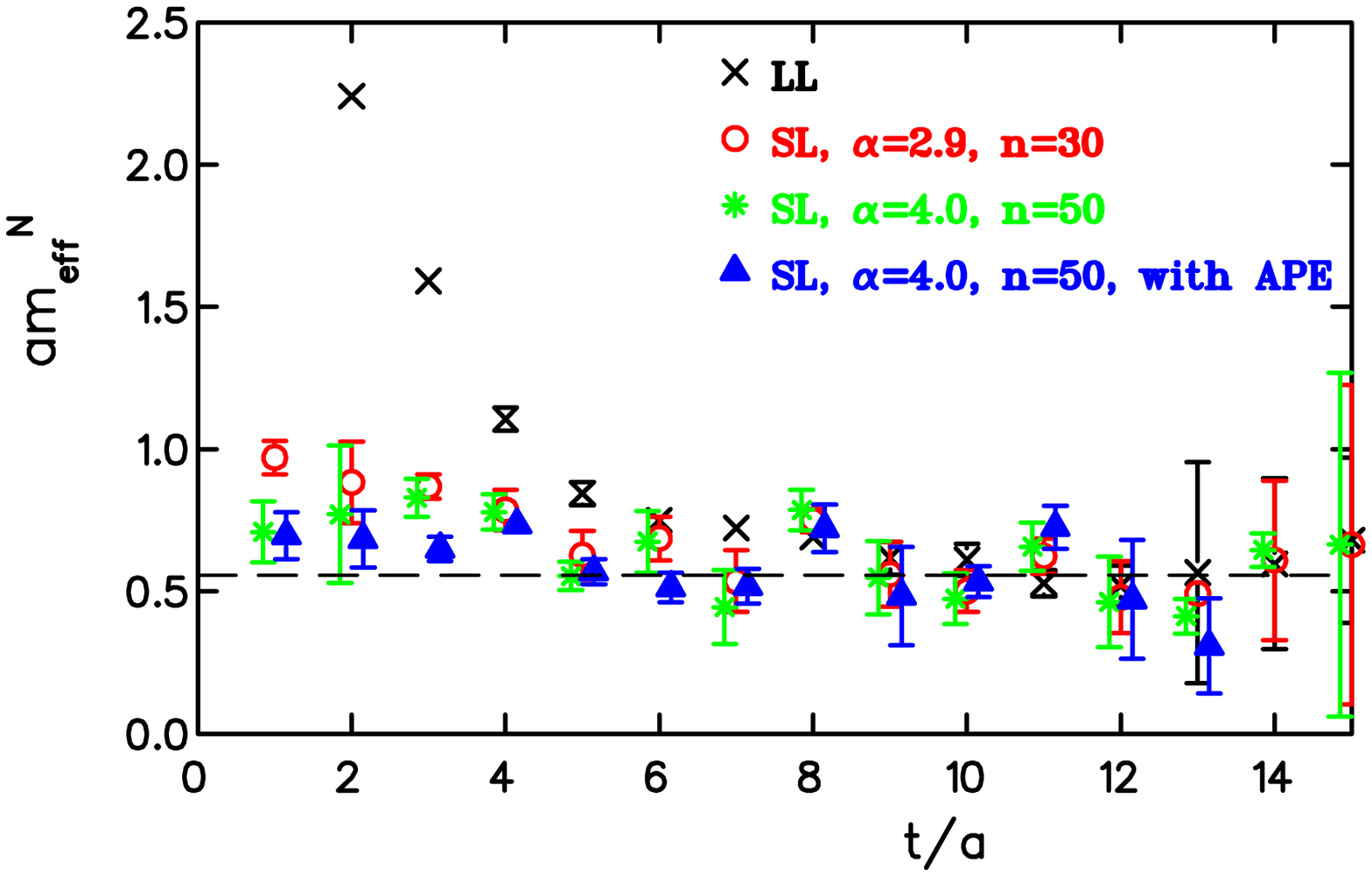}}
\caption{
$m_{\rm eff}^N$ versus time separation both in
lattice units. Crosses show
results using local sink and source (LL), 
circles (asterisks) using Gaussian smearing at the sink (SL) with
 $\alpha=2.9$ and $n=30$ ( $\alpha=4$ and $n=50$),
  and filled triangles with $\alpha=4$ and $n=50$ 
 and APE smearing. The dashed line is the plateau value extracted
by fitting results when APE smearing
is used.}
\label{fig:fit ratio}
\label{fig:meff_opt}
\end{minipage}
\end{figure}

\noindent
 Local interpolating fields 
 are not
optimal for suppressing excited state contributions. We  apply
 Gaussian
 smearing to each  quark field,  $q({\bf x},t)$~\cite{Gusken:1989,Alexandrou:1992ti}. The smeared quark field
is given by 
$q^{\rm smear}({\bf x},t) = \sum_{\bf y} F({\bf x},{\bf y};U(t)) q({\bf y},t)$
using the gauge invariant smearing function
\be 
F({\bf x},{\bf y};U(t)) = (1+\alpha H)^ n({\bf x},{\bf y};U(t)),
\ee
constructed from the
hopping matrix understood as a matrix in coordinate, color and spin space,
\be
H({\bf x},{\bf y};U(t))= \sum_{i=1}^3 \biggl( U_i({\bf x},t)\delta_{{\bf x,y}-a\hat \imath} +  U_i^\dagger({\bf x}-a\hat \imath,t)\delta_{{\bf x,y}+a\hat \imath}\biggr).
\ee
The parameters $\alpha$ and $n$ are varied so 
that the root mean square (r.m.s.)  radius  obtained
 using the proton interpolating field is in the range of 0.3-0.4~fm.
In Fig.~\ref{fig:contour} we show lines of constant r.m.s radius as
we vary $\alpha$ and $n$.
The larger the  $n$ the more time consuming is the smearing procedure.
 On the other hand,
for  $\alpha\stackrel{>}{\sim}1$, increasing further $\alpha$ 
does not reduce $n$ significantly. 
Therefore, we choose a value of $\alpha$ large enough 
so that the weak $\alpha$-dependence sets in, and we
adjust $n$ to obtain the required value of the r.m.s radius.
We consider two sets for these parameters  
giving r.m.s radius 0.32~fm and 0.41~fm, as  
shown in Fig.~\ref{fig:contour}. For each set of parameters
we evaluate the effective mass as 
\be
m_{\rm eff}^P=-\log(C_P(t)/C_P(t-1))
\ee
where $C_P(t)$ is the zero-momentum 
two-point correlator of the particle $P$ given by
\bea
C_P(t) = \frac 1 2 {\rm Tr}(1 \pm \gamma_4) \sum_{\bf x_{\rm sink}}
\langle J_P( {\bf x}_{\rm sink}, t_{\rm sink}) \bar J_P({\bf x}_{\rm source}, t_{\rm source})\rangle,\qquad 
t=t_{\rm sink}-t_{\rm source} \quad.
\label{C_P}\eea
In Fig.~\ref{fig:meff_opt}, we show the nucleon effective mass,
 $m_{\rm eff}^N$ in lattice units
  for 10 configurations  at $\beta=3.9$
and $a\mu=0.0085$. For the optimization of the parameters 
we apply Gaussian smearing at the
sink, whereas for  the source we use local interpolating fields
so that no additional inversions are needed when we change $\alpha$ and $n$.
As can be seen, for both sets of smearing parameters,
the excited state contributions are suppressed with the set $\alpha=4$, $n=50$
producing  a plateau a couple of time slices earlier.
 If, in addition, we apply APE smearing~\cite{Albanese:1987}
 to the spatial links 
that enter the hopping matrix in the smearing function,
then  gauge noise  is reduced resulting in a better identification of the 
plateau.
Therefore for all computations at  $\beta=3.9$ we  use Gaussian smearing with 
$\alpha=4$ and $n=50$. 
Having chosen the smearing parameters, for the results that follow,
 we apply smearing at the source
and compute the mass using both local (LS) and smeared sink (SS).
For $\beta=4.05$ we readjust the parameters so that the nucleon 
r.m.s radius is still about 0.4~fm, obtaining  $\alpha=4$ and $n=70$.
For $\beta=3.8$ we use $\alpha=4$ and $30$ to keep the r.m.s. radius at the
 same value.
Also for these two values of $\beta$
 we apply APE smearing to the gauge links that
are used in $F({\bf x},{\bf y};U(t))$.

There are other methods to enhance ground state dominance
besides  Gaussian smearing.  Smearing based on link 
  fuzzing has been effectively  used in the pion sector.
 In this work, having optimized 
our parameters for Gaussian smearing
  we  use only  local and Gaussian-smeared 
  interpolating fields.


\subsection{Two point correlators}

The lowest energy state with which the nucleon interpolating field
given in ~\eq{interpolate}
has a non-vanishing overlap is the proton state $|p({\bf p},s)\rangle$
\be
   \langle 0| J_p | p({\bf p}, s) \rangle = Z_p u({\bf p}, s) \, .
   \label{nucleonoverlap}
\ee
$Z_p$ is a constant overlap factor and $u({\bf p}, s)$, with $s \in \{-1/2, +1/2\}$, is a
solution of the Dirac equation
\be
   (\pslash - m_N) u = 0 \, .
\ee
Averaging over the nucleon spins and choosing the nucleon rest frame, we 
are led to the two point correlator
\be
   C_N^\pm(t) = \frac 1 2 {\rm Tr}(1 \pm \gamma_4) \sum_{{\bf x}_{\rm sink}}
   \langle J_N( {\bf x}_{\rm sink}, t_{\rm sink}) \bar{J}_N({\bf x}_{\rm source}, t_{\rm source})\rangle,\qquad 
   t=t_{\rm sink}-t_{\rm source} \, .
   \label{C_N}
\ee
Space-time reflection symmetries of the action and the antiperiodic boundary conditions in 
the temporal direction for the quark fields
imply, for zero three-momentum correlators, 
that $C_N^+(t) = -C_N^-(T-t)$. The nucleon mass is
extracted from the exponential decay of the correlator
\be
   C_N(t) = C_N^+(t) - C_N^-(T-t) \, .
\ee 
To increase the precision we 
also average over the proton and neutron correlators since these 
are degenerate in mass.

\begin{figure}[htbp]
\begin{minipage}{8cm}
\mbox{\epsfxsize=8.cm\epsfysize=6.cm\epsffile{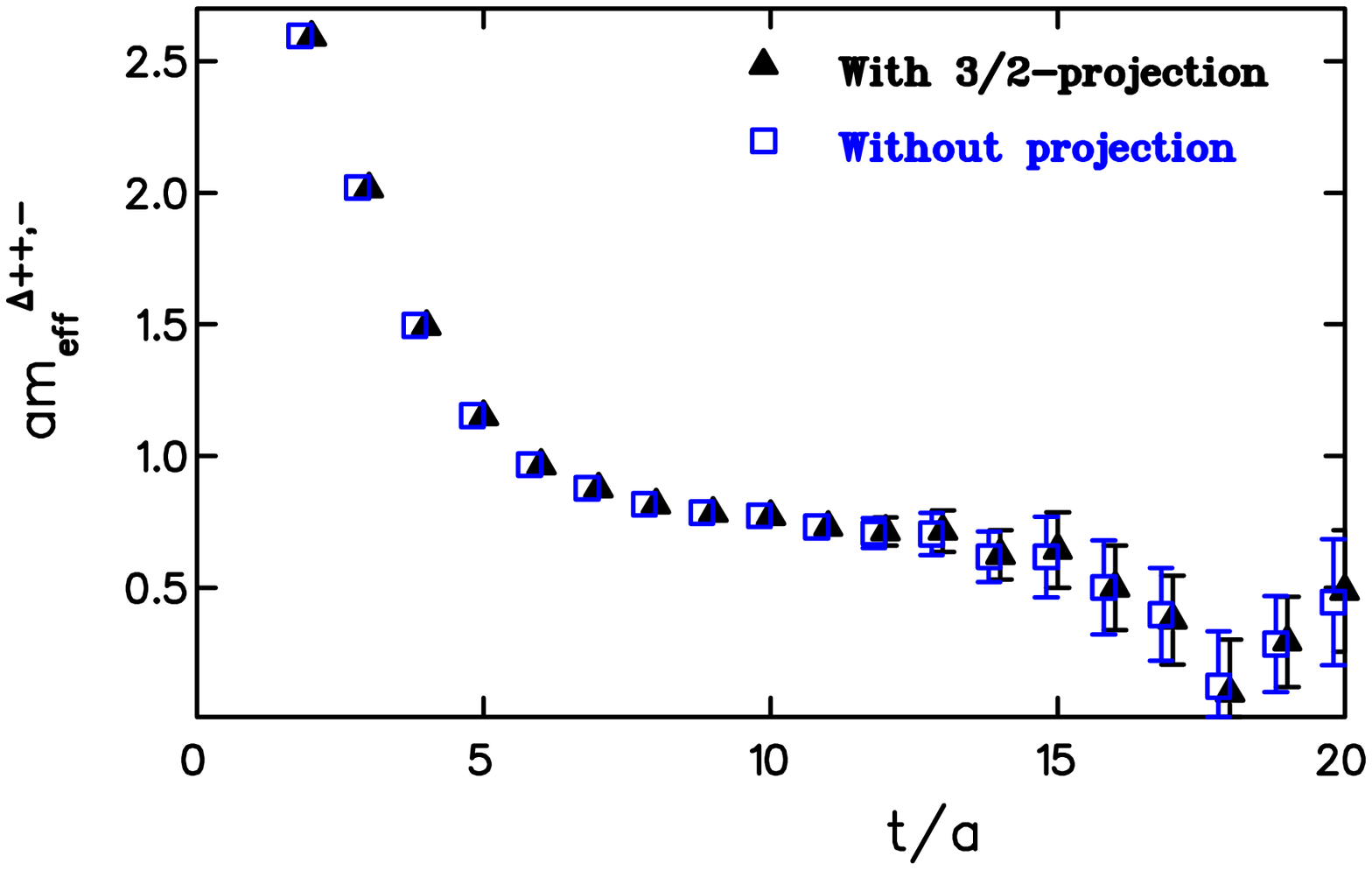}}
\caption{ Comparison of effective masses for $\Delta^{++,-}$ for
$a\mu=0.0085$ at $\beta=3.9$ on the lattice volume $24^3 \times48$, 
obtained with (filled triangles) or without (open squares shifted to the 
left for clarity) spin projection,
using a sample of 90 configurations. 
The mass difference
with projection and without projection 
is much smaller
than the statistical error.}
\label{fig:comparasion}
\end{minipage}
\hfill
\begin{minipage}{8cm}
\mbox{\epsfxsize=8.cm\epsfysize=6.cm\epsffile{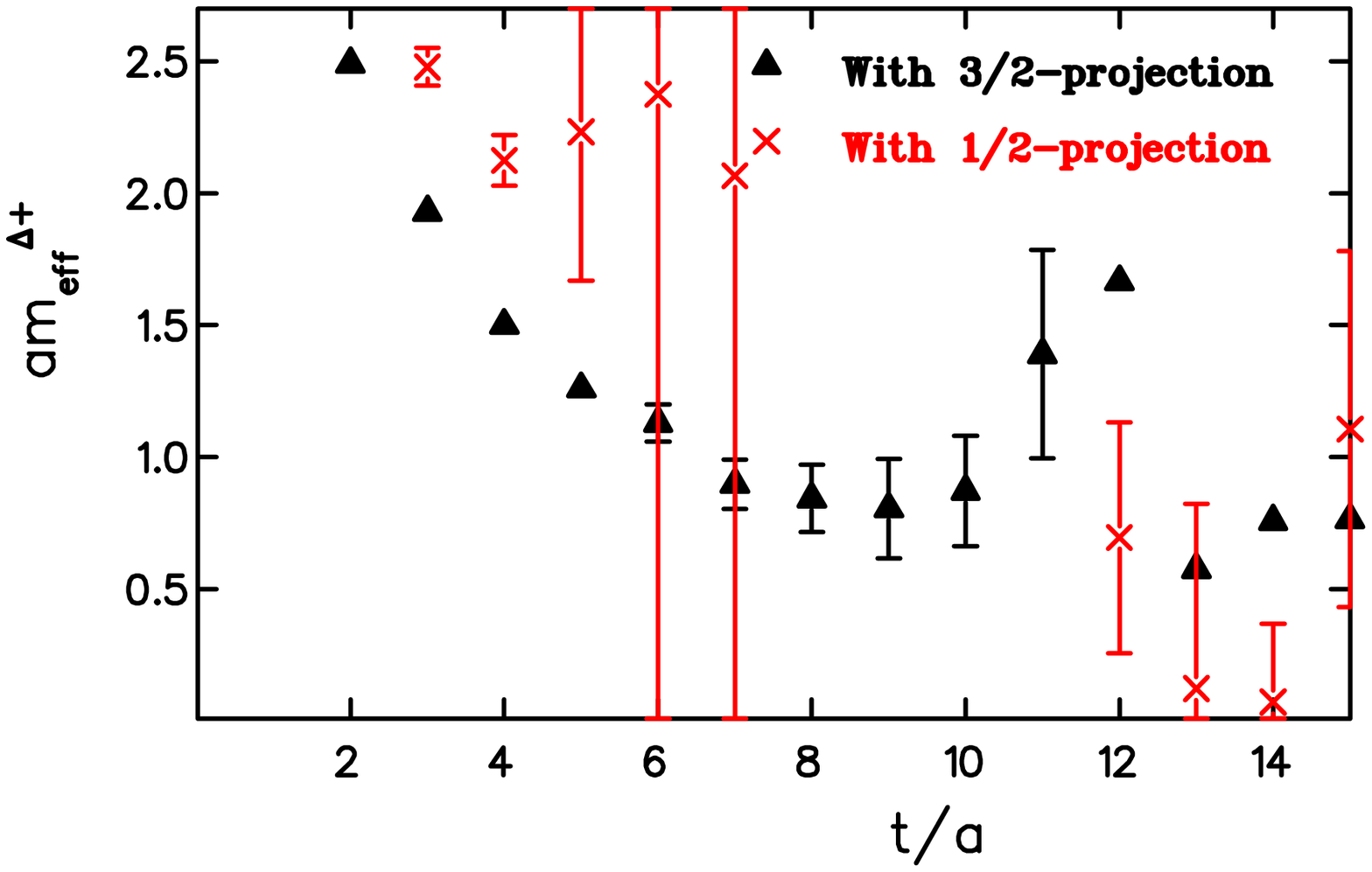}}
\caption{ Comparison of effective masses for $\Delta^{+}$ for
$a\mu=0.011$ at $\beta=3.8$ on the lattice volume $24^3 \times48$, 
obtained with $3/2$-spin (filled triangles) or with
$1/2$-spin  projection,
using a sample of 50 configurations. 
}
\label{fig:spin12and23}
\end{minipage}
\end{figure}

In analogy to~\eq{nucleonoverlap}, the overlap of the $\Delta^+$ interpolating field with the 
$\Delta^+$ state is given by
\be
   \langle 0|J^\mu_{\Delta^+} | \Delta^+({\bf p}, s) \rangle = Z_{\Delta^+} u^\mu({\bf p}, s) \, .
\ee
Every  vector component of the Rarita-Schwinger spinor $u^\mu$ satisfies the Dirac equation
\be
   (\pslash - m_\Delta) u^\mu = 0 \qquad \mu=1\ldots 4 \, ,
\ee
and in addition the auxiliary conditions
\be\label{auxiliarycond}
   p_\mu u^\mu =0 \qquad {\rm and} \qquad \gamma_\mu u^\mu = 0
\ee
are fulfilled. The four independent solutions are labeled by
$s\in \{-3/2, -1/2, 1/2, 3/2\}$. 
The $\Delta$ interpolating fields 
as defined in~\eq{interpolate}
have overlap also with the heavier spin-1/2 excitations. These overlaps 
can be removed when  the conditions in ~\eq{auxiliarycond} are 
enforced on the interpolating fields. This can be achieved by
the incorporation of a spin-3/2 projector in the 
definitions of the interpolating fields
\bea
   J^\mu_{3/2} &=&  P_{3/2}^{\mu\nu} J_{\nu \Delta} \\
   P_{3/2}^{\mu\nu} &=& \delta^{\mu\nu} -\frac{1}{3}\gamma^\mu\gamma^\nu - \frac{1}{3p^2}(\pslash\gamma^\mu p^\nu + p^\mu \gamma^\nu \pslash) \, .
\eea
Similarly the spin $1/2$-interpolating field, $J^\mu_{1/2}$,
that has only overlap with the $1/2$-state,
is obtained by acting with the spin $1/2$-projector 
$P_{1/2}^{\mu\nu}=g^{\mu\nu}- P_{3/2}^{\mu\nu}$ on $J_\Delta^\mu$.
Using any of the three interpolating fields, the $\Delta$ masses
are extracted from the two-point functions
\bea
C^\pm_\Delta(t) = \frac 1 6 {\rm Tr}(1 \pm \gamma_4) 
\sum_{{\bf x}_{\rm sink}}\sum_{i=1}^3
\langle J^i_\Delta({\bf x}_{\rm sink},t_{\rm sink}) 
\bar{J}^i_\Delta({\bf x}_{\rm source},t_{\rm source})\rangle,\qquad 
t=t_{\rm sink}-t_{\rm source}.
\eea\label{C_Delta}
Fig.~\ref{fig:comparasion} compares effective masses extracted from
correlation functions with and without the spin~3/2 projection
at $\beta=3.9$. For this comparison we
use 90 configurations, a number sufficient for the purpose of this check. 
The results for the effective mass are hardly affected by including the
spin-3/2 projector even at very short time separations. This is because
the overlap of the interpolating field $J_\Delta^\mu$ with the spin-1/2 state
is small, a property that holds at all values of $\beta$ 
This is clearly seen in Fig.~\ref{fig:spin12and23} at $\beta=3.8$ where
the effective mass obtained using the spin-1/2 projected interpolating field
$J_{1/2}^\mu$ is much more noisy than with $J_{3/2}^\mu$ due to the small
overlap with the spin-1/2 state. This behavior is in agreement
 with the results of Ref.~\cite{Zanotti:2003} where the same spin projections
were implemented.
Since the impact on the plateau value is negligible compared to the
statistical uncertainty, we use only the non-projected interpolating 
fields from here on. We average the correlators of
$\Delta^{++}$ with $\Delta^{-}$ as well as $\Delta^{+}$ with $\Delta^{0}$.
We do not average the $\Delta^{++}$ and $\Delta^{+}$ 
correlators as they differ by an 
$O(a^2)$ isospin breaking effect.

\subsection{Effective masses}


\begin{figure}[h]
\begin{minipage}{8cm}
\epsfxsize=8truecm
\epsfysize=10truecm
 \mbox{\epsfbox{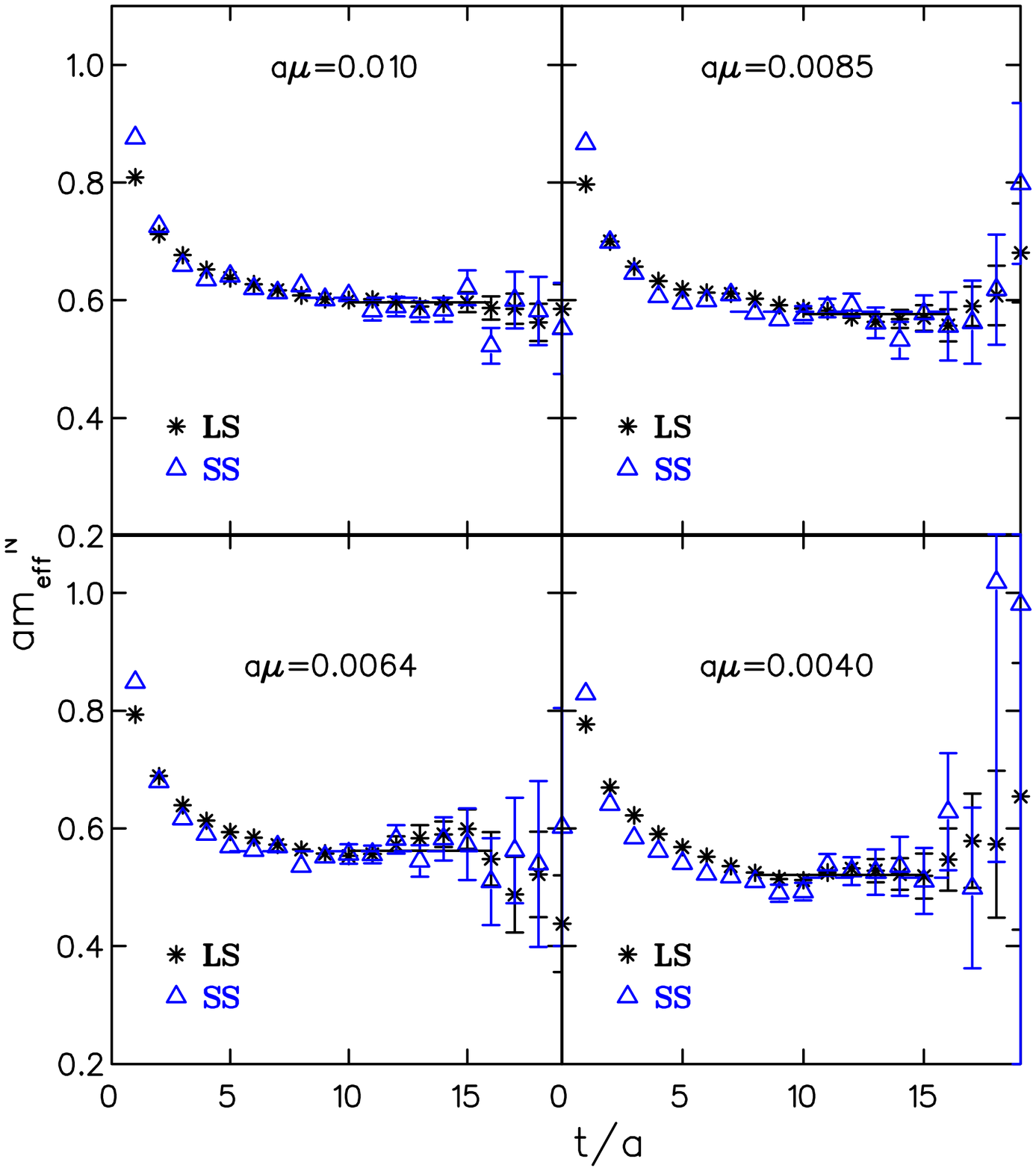}}
\caption{Nucleon effective mass  (LS: asterisks, SS: open triangles)
for $\beta=3.9$  versus
time separation in lattice units,
for
$a\mu=0.010$ (upper left), $0.0085$ (upper right), $0.0064$ (lower left) 
and $0.0040$ (lower right).
The constant lines
are the best fits to the data over the range spanned by the lines. }
\label{fig:nucleon meff}
\end{minipage}
\hfill
\begin{minipage}{8cm}
\epsfxsize=8truecm
\epsfysize=10truecm
 \mbox{\epsfbox{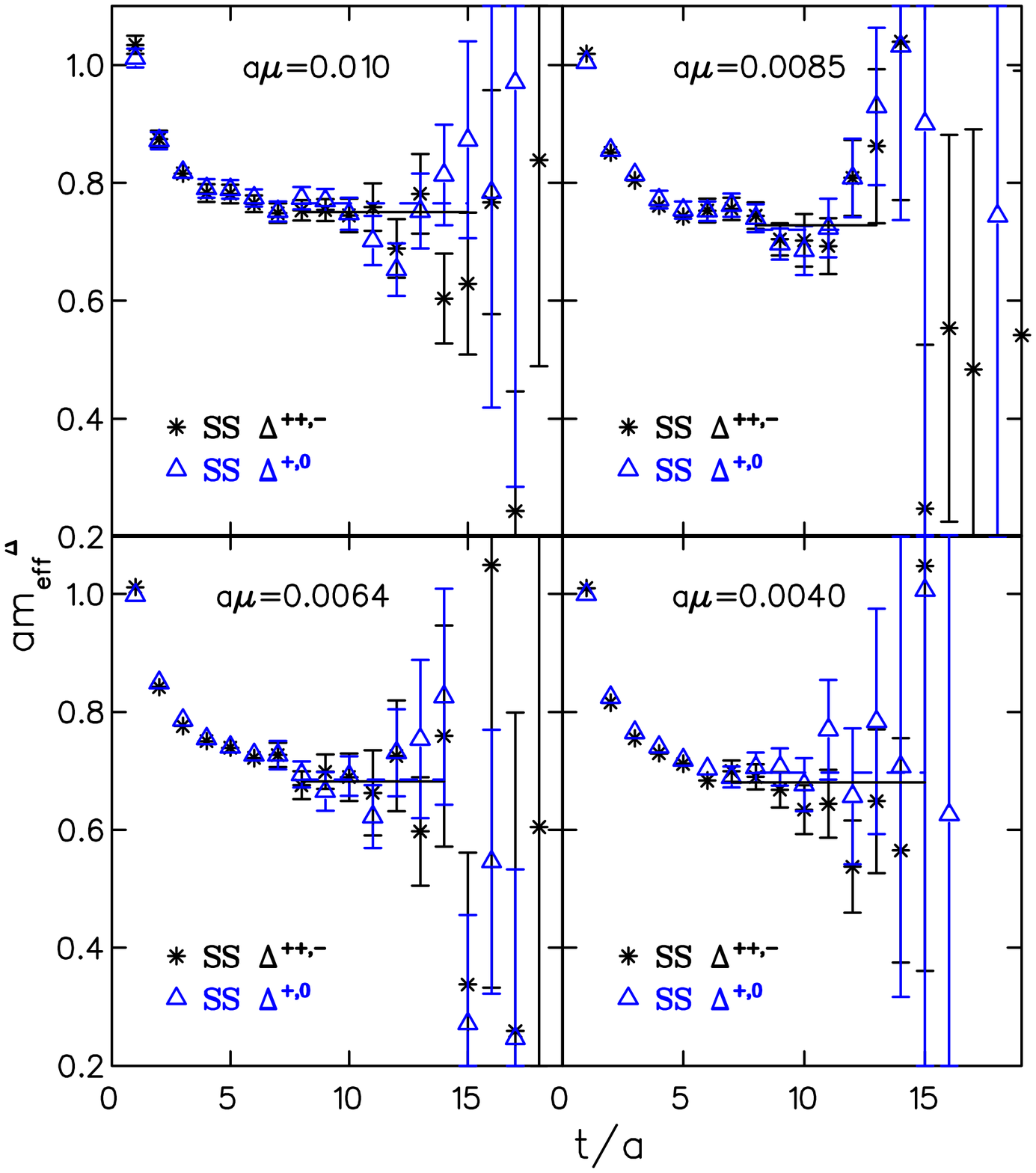}}
\caption{$\Delta^{++,-}$ (asterisks) and $\Delta^{+,0}$ (open triangles)
 effective masses using smeared-smeared (SS) correlators for $\beta=3.9$ %
versus
time separation in lattice units, for
$a\mu=0.010$ (upper left), $0.0085$ (upper right), $0.0064$ (lower left) 
and $0.0040$ (lower right).
The straight lines
are the best fits to the data over the range spanned by the lines.
The solid line is for $\Delta^{++,-}$ and the dashed for $\Delta^{+,0}$
and they coincide. }
\label{fig:delta meff}
\end{minipage}
\end{figure}


\begin{table}[h]
  
  \begin{tabular}{lcccccc}
    \hline\hline
    $\Bigl.\Bigr.a\mu$ & Interpolating field
    & number of confs.    & $am_{\pi}$  & $am_N$ & $am_{\Delta^{++,-}}$ &
    $am_{\Delta^{+,0}}$  \\
    \hline\hline
\multicolumn{7}{c}{$24^3\times 48$}\\
\hline
   $0.0040$  &LL&   471   &  $0.13587(68)$ & $0.511(11)$ & $0.699(15)$ & $0.708(25)$ \\
   $0.0040$  &LS& 419 &  $0.13587(68)$ & $0.521(6) $ & $0.694(11)$ & $0.717(13)$ \\
   $0.0040$  &SS& 419 &  $0.13587(68)$ & $0.515(5) $ & $0.682(12)$ & $0.697(17)$ \\
   $0.0064$  &LL&  199   &  $0.16937(36)$ & $0.565(11)$ & $0.727(15)$ & $0.763(15)$ \\
   $0.0064$  &LS& 235 &  $0.16937(36)$ & $0.565(6) $ & $0.715(13)$ & $0.742(7)$ \\
   $0.0064$  &SS& 235 &  $0.16937(36)$ & $0.561(4) $ & $0.710(11)$ & $0.711(10)$ \\
   $0.0085$  &LL&  153    &  $0.19403(50)$ & $0.568(8) $ & $0.754(12)$ & $0.776(22)$\\
   $0.0085$  &LS& 186 &  $0.19403(50)$ & $0.581(6) $ & $0.746(9) $ & $0.751(12)$ \\
   $0.0085$  &SS& 186 &  $0.19403(50)$ & $0.580(6) $ & $0.738(11)$ & $0.742(9)$ \\
   $0.0100$  &LL& 173     &  $0.21004(52)$ & $0.613(6) $ & $0.823(7)$* & $0.767(12)$\\
   $0.0100$  &LS& 213 &  $0.21004(52)$ & $0.595(7) $ & $0.742(7) $ & $0.760(7)$\\
   $0.0100$  &SS& 213 &  $0.21004(52)$ & $0.589(9) $ & $0.750(10)$ & $0.755(10)$\\
\hline\hline
\multicolumn{7}{c}{$32^3 \times 64$}\\
    \hline
   $0.0040$  &LS& 201 &  $0.13377(24)$ & $0.518(5) $ & $0.672(9) $ & $0.670(14)$ \\
   $0.0040$  &SS& 201 &  $0.13377(24)$ & $0.510(5) $ & $0.660(9) $ & $0.660(14)$ \\
      \hline\hline
  \end{tabular}
  \caption{Results for the nucleon and $\Delta$ mass 
 at $\beta=3.9$ 
for  lattices of size  $24^3\times 48$ and  $32^3\times 64$.
LL stands for local sink and local source, LS for local sink and
smeared source and SS for smeared sink and smeared source.
    The results for the pion mass are taken 
    from Table~2 of Ref.~\cite{Boucaud:2006} computed using
more gauge configurations than we used for the evaluation of the baryon
masses as well as a different smearing and therefore are the same
for LL, LS and SS.
With an asterisk we mark results  for which the effective
mass  does not show a good plateau.    } 
  \label{tab:masses-3.9}
\end{table}
%
%
\begin{table}[h]
  
  \begin{tabular}{lcccccc}
    \hline\hline
    $\Bigl.\Bigr.a\mu$ & Interpolating field & number of confs.
       & $am_\pi$ & $am_N$ & $am_{\Delta^{++,-}}$ &
    $am_{\Delta^{+,0}}$  \\
    \hline\hline
    $0.0030$ & LL&  70     & $0.1038(6)$ & $0.403(15)$ & $0.633(30)$ &  \\
    $0.0030$ & LS& 201 & $0.1038(6)$ & $0.396(7)$  & $0.536(18)$ & $0.546(12)$ \\
    $0.0030$ & SS& 201 & $0.1038(6)$ & $0.402(8)$  & $0.538(19)$ & $0.536(15)$ \\
    $0.0060$ & LL&  216    & $0.1432(6)$ & $0.453(5)$  & $0.597(8) $ & $0.575(9)$ \\
    $0.0060$ & LS& 160 & $0.1432(6)$ & $0.448(5)$  & $0.564(7) $ & $0.566(7)$ \\
    $0.0060$ & SS& 160 & $0.1432(6)$ & $0.446(6)$  & $0.562(6) $ & $0.566(7)$ \\
    $0.0080$ & LL&  240    & $0.1651(5)$ & $0.465(6)$  & $0.627(6) $ & $0.637(7)$\\
    $0.0080$ & LS& 256 & $0.1651(5)$ & $0.469(4)$  & $0.590(7) $ & $0.585(9)$ \\
    $0.0080$ & SS& 256 & $0.1651(5)$ & $0.465(5)$  & $0.594(7) $ & $0.594(8)$ \\
    $0.0120$ & LL& 157    & $0.2025(6)$ & $0.520(5)$  & $0.670(4)$* & $0.677(5)$\\
    $0.0120$ & LS& 162 & $0.2025(6)$ & $0.509(4)$  & $0.616(7)$* & $0.623(7)$\\
    $0.0120$ & SS& 162 & $0.2025(6)$ & $0.515(3)$  & $0.616(7)$* & $0.620(7)$\\
    \hline\hline
  \end{tabular}
  \caption{Results for the nucleon and $\Delta$ mass 
 at $\beta=4.05$ for the $32^3\times 64$ lattice. 
LL stands for local sink and local source, LS for local sink and
smeared source and SS for smeared sink and smeared source.
The results for the pion mass are computed using
more gauge configurations than we used for the evaluation of the baryon
masses as well as a different smearing~\cite{ETMC:future} 
and therefore are the same
for LL, LS and SS. .
With an asterisk we mark results  for which the effective
mass  does not show a good plateau.  
Empty entries are due to the absence of a sufficient 
plateau region. }  
  \label{tab:masses-4.05}
\end{table} 

\begin{table}[h]
  
  \begin{tabular}{ccccccc}
    \hline\hline
    $\Bigl.\Bigr.a\mu$ & Interpolating field & number of confs.
        & $am_\pi$ & $am_N$ & $am_{\Delta^{++,-}}$ &
    $am_{\Delta^{+,0}}$  \\
    \hline\hline
    $0.0060$ & LL & 211 & $0.1852(10)$ & $0.623(20)$ & $0.792(25)$ & $0.815(28)$ \\
    $0.0060$ & SL & 211 & $0.1852(10)$ & $0.637(9) $ & $0.818(11)$ & $0.824(13)$ \\
    $0.0080$ & LL & 283 & $0.2085(8)  $ & $0.676(11)$ & $0.859(11)$ & $0.847(30)$ \\
    $0.0080$ & SL & 283 & $0.2085(8)  $ & $0.665(9) $ & $0.827(17)$ & $0.856(24)$ \\
    $0.0110$ & LL & 251 & $0.2424(5)  $ & $0.700(9) $ & $0.861(13)$ & $0.893(22)$ \\ 
    $0.0110$ & SL & 251 & $0.2424(5)  $ & $0.699(8)$  & $0.854(14)$ & $0.875(16)$  \\
    $0.0165$ & LL & 249 & $0.2957(5)  $ & $0.759(7) $ & $0.948(12)$ & $0.886(25)$ \\
    $0.0165$ & SL & 249 & $0.2957(5)  $ & $0.744(8) $ & $0.942(13)$ & $0.946(12)$ \\
    \hline\hline
  \end{tabular}
  \caption{
Results for the nucleon and $\Delta$ mass 
 at $\beta=3.8$ for the $24^3\times 48$ lattice. 
  The notation is the same as that of Table~ \ref{tab:masses-4.05} with SL
being a smeared sink and local source. 
   } 
  \label{tab:masses-3.8}
\end{table}
 In Fig.~\ref{fig:nucleon meff} we show the nucleon effective masses
   at $\beta=3.9$ on a
 volume $24^3\times 48$ for all the values of $\mu$ considered.
We smeared the  source as described in the previous section
and either use a local sink or smear the sink with the same smearing
used for the source. As expected, the effective masses are consistent
for both smeared and local sink yielding asymptotically the same constant.
We fit the effective mass to a constant in the region where $m_{\rm eff}(t)$
becomes time independent (plateau region) and vary the lower t-range of
the fit so that $\chi^2$ per degree of freedom (d.o.f.) becomes
less than one. We take this value for the mass of the nucleon.        
 In Fig.~\ref{fig:delta meff} we show,  for the same $\mu$-values, 
effective masses for the $\Delta^{++,-}$ and $\Delta^{+,0}$ 
 using smeared source and sink.
We fit in the same way as in the nucleon case to extract the mass of the $\Delta$.
As can be seen,
 the quality of the plateaus in the nucleon
case is better than in the case of the $\Delta$. This explains why results
on the $\Delta$ mass have larger errors.
The errors are evaluated using jackknife and
the $\Gamma$-method~\cite{Wolff:2004} to check consistency.
The integrated auto-correlation times for our baryonic observables 
are very small for our configuration ensembles. Since for our 
computation  we use gauge configurations that  are separated 
by 8-20 trajectory lengths, autocorrelations are negligible.

The  resulting masses using local and smeared 
interpolating fields 
are summarized   in Table \ref{tab:masses-3.9} for $\beta=3.9$ 
using lattice sizes of $24^3\times 48$  
and $32^3\times 64$, while
 those obtained for $\beta=4.05$ on a lattice volume of $32^3 \times 64$
are reported in Table \ref{tab:masses-4.05}. Results obtained at $\beta=3.8$
are given in Table~\ref{tab:masses-3.8}.
The mass of the pion listed in Table~\ref{tab:masses-3.9}
 is taken from Ref.~\cite{Boucaud:2006}
and was evaluated using a larger set
of configurations applying a different smearing than the one used in this work. A detailed description of this evaluation as well as the error analysis
is presented in Ref.~\cite{ETMClong}.
The pion masses given in Tables~\ref{tab:masses-4.05} and \ref{tab:masses-3.8}
are again obtained in a separate evaluation~\cite{ETMC:future}.

\section{lattice artifacts}
\subsection{Finite volume effects}

 At $\beta=3.9$ and for $a\mu =0.004$ we have gauge configurations on two lattices of different volume.
This is the smallest $\mu$-value considered at $\beta=3.9$ 
and it is the one that potentially can have the largest finite volume effects. 
On the lattice of spatial extension $L_s=24$  the other three
larger $\mu$-values satisfy
 the condition  $m_\pi L_s \ge 4$, 
whereas for $a\mu=0.004$
we have $m_\pi L_s \sim 3.2$. On the $32^3$ lattice at $a\mu=0.004$ we have 
$m_\pi L_s > 4$.
Applying the
re-summed L\"uscher formula to the nucleon mass and using 
 the knowledge of the $\pi N$ scattering amplitude to ${\cal O}(p^2)$
and  ${\cal O}(p^4)$, 
the volume corrections are estimated  to be about 
3\% to 5\%~\cite{Colangelo:2006} for $L_s\sim 2$~fm and $m_\pi\sim 300$~MeV. 
In Table~\ref{tab:masses-3.9} we give the results
for the nucleon mass using our two lattice volumes. The smaller lattice volume  has spatial length 
very close to  the  2~fm  length of Ref.~\cite{Colangelo:2006} namely
$L_s\sim 2.1$~fm. The results for $m_N$ do not
 change within our statistical accuracy
when we use  the larger lattice size of $L_s\sim 2.7$~fm.
We make the  assumption that for the larger lattice finite volume
corrections have become  negligible and take them to be
a good approximation to the infinite volume
 results.
 In other words we take $m_N(L_s=\infty)\simeq m_N(L_s=2.7\; {\rm fm})$.
This assumption was shown to be valid in the pion sector where
 a finite size analysis was carried out~\cite{ETMClong}. 
We define the ratio 
 $R_N\equiv\Delta m_N/m_N(L_s=\infty)$, where  
$\Delta m_N\equiv m_N(L_s=2.1 \;{\rm fm})-m_N(L_s=\infty)$ 
and  estimate $R_N$ with
results obtained on our two volumes for the smallest pion mass. This
gives us an estimate for our finite volume errors. 
Using the results tabulated in  Table~\ref{tab:masses-3.9}
at $a\mu=0.004$ we conclude that
 $\Delta m_N$ is positive as expected.
This is also true for the corresponding difference for the masses
 of $\Delta^{++,-}$ and $\Delta^{+,0}$. In Table~\ref{tab:delta_m} we
give the ratios $R_N$, $R_{\Delta^{++,-}}$ and    $R_{\Delta^{+,0}}$.
 For the nucleon this ratio is compatible with zero
and within our accuracy it can be
at the most 2\%. For the $\Delta^{++,-}$ where the statistical 
errors are smaller than for the  $\Delta^{+,0}$, the volume corrections range
from 1\% to 5\%. From this study we conclude that finite volume effects
on the nucleon mass are negligible whereas for the $\Delta$ we can at most
have corrections on the few percent level.

\begin{table}[h]
  
  \begin{tabular}{lcccc}
     \hline\hline
    $\Bigl.\Bigr.a\mu$ & Interpolating field 
        & $R_N$ & $R_{\Delta^{++,-}}$ &
    $R_{\Delta^{+,0}}$  \\
    \hline\hline
          $0.0040$ & LS&  $0.006(15)$ & $0.033(22)$ & $0.070(26)$ \\
          $0.0040$ & SS&  $0.010(14)$ & $0.033(23)$ & $0.056(33)$ \\
      \hline\hline
  \end{tabular}
  \caption{Finite volume dependence 
    at $\beta=3.9$ for $a\mu=0.004$. Results with a lattice of size 
     $24^3 \times 48$ are compared to those obtained with a lattice size of $32^3\times 64$. 
  For a hadron state $P$  we define  $R_P \equiv     (m_P(L_s=2.1\;{\rm fm})-m_P(L_s=2.7\;{\rm fm}))/m_P(L_s=2.7\;{\rm fm}) 
    \simeq   (m_P(L_s=2.1\;{\rm fm})-m_P(\infty))/m_P(\infty)$
 assuming the masses at
    $2.7$ fm to be close enough to the infinite volume limit.
  }
  \label{tab:delta_m}
  \end{table} 
 
\subsection{Isospin breaking}


\begin{figure}[h]
\begin{minipage}{8cm}
\epsfxsize=8truecm
\epsfysize=8truecm
 \mbox{\epsfbox{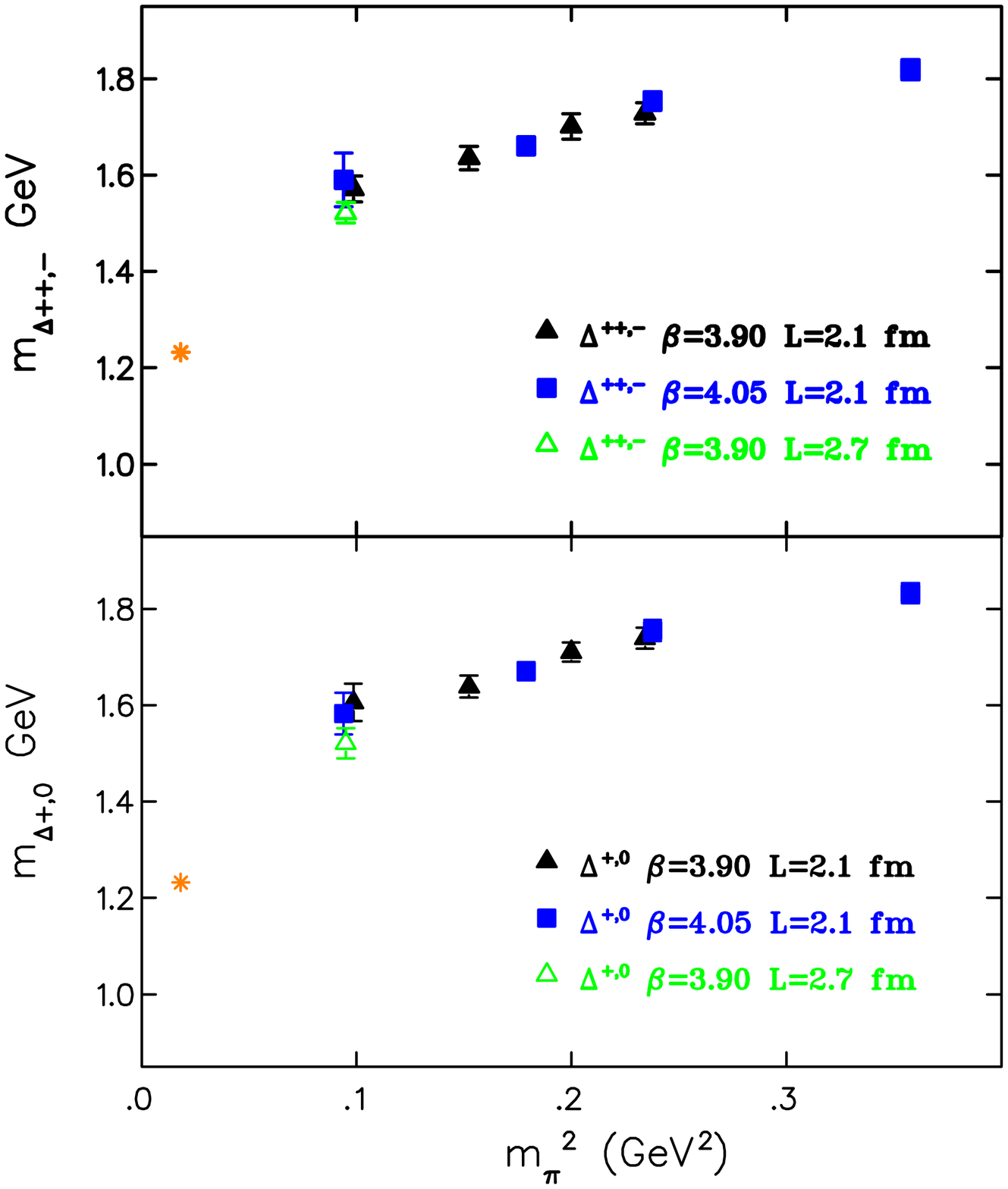}}
\caption{The $\Delta^{++,-}$  (upper graph) 
and $\Delta^{+,0}$ (lower graph) mass as a function of $m_\pi^2$
for $\beta=3.9$ on a lattice of size $24^3\times 48$ (filled triangles)
and on a lattice of size $32^3\times 64$ (open triangles). Results
at $\beta=4.05$ are shown with the filled squares. The physical $\Delta$
mass is shown with the asterisk.}
\label{fig:md}
\end{minipage}
\hfill
\begin{minipage}{8cm}
\epsfxsize=8truecm
\epsfysize=6truecm
 \mbox{\epsfbox{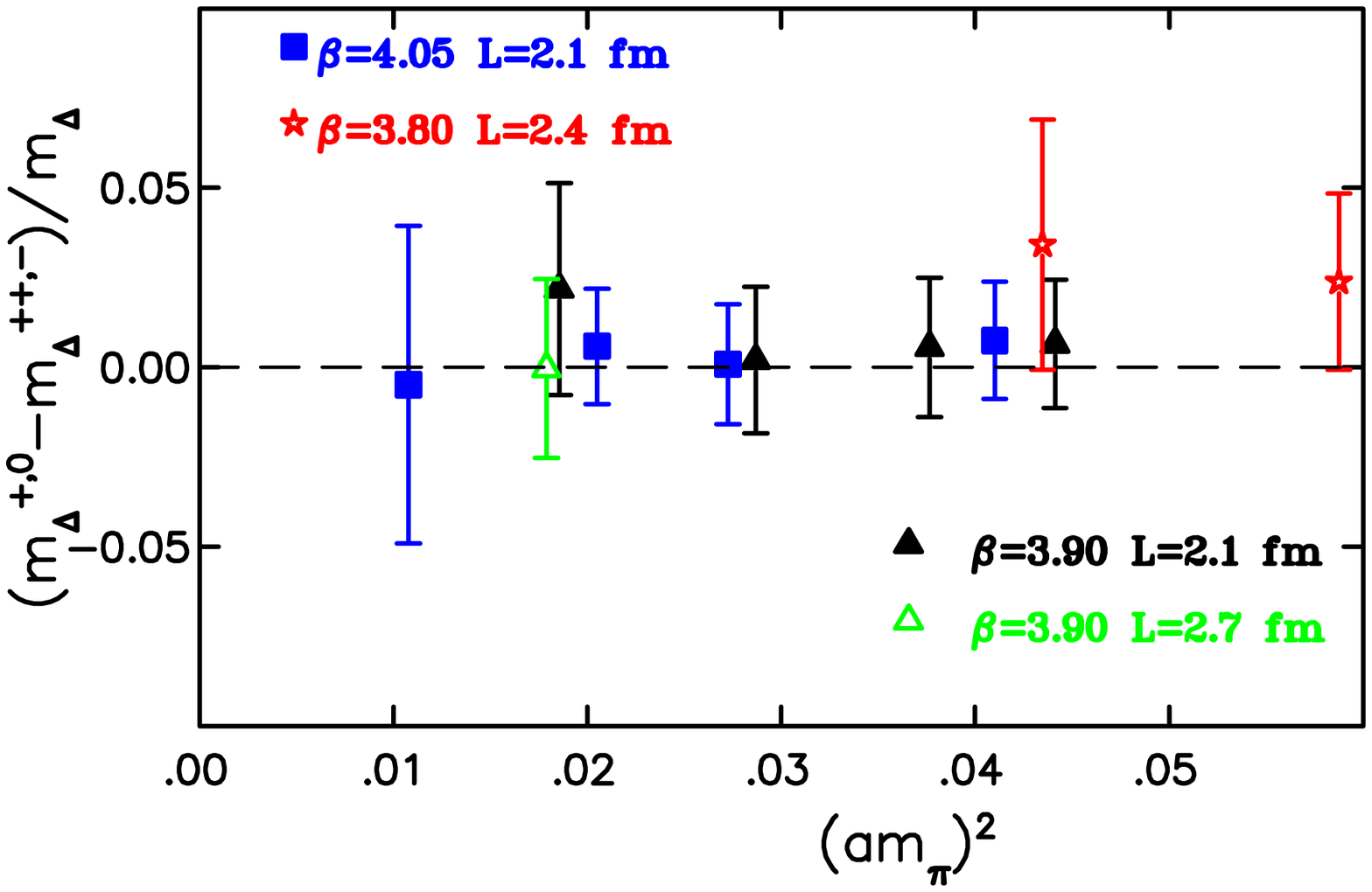}}
\caption{The mass 
 splitting between $\Delta^{+,0}$ and $\Delta^{++,-}$ normalized with the mean value of their mass $m_\Delta$
  as a function of $m_\pi^2$  in lattice units. Results
at $\beta=3.8$ are shown with the asterisks. The rest of the notation is the
same as in Fig.~\ref{fig:md}.}
\label{fig:Dm}
\end{minipage}
\end{figure}

One of the main goals of this work is to examine isospin breaking in
the baryon sector due to lattice artifacts.
 As already explained  the proton and the neutron
 are degenerate. Isospin breaking in the light baryon sector
can be examined  for the $\Delta$.
In Fig.~\ref{fig:md} we show results for the mass of $\Delta^{++,-}$
 as well as for the mass of $\Delta^{+,0}$.  Results at
 $\beta=3.9$ and $\beta=4.05$ fall on the same curve pointing to 
small cut-off effects. Small  finite volume effects are visible
at the smallest pion mass at $\beta=3.9$ as discussed in the previous
subsection. To check for isospin breaking we plot the mass difference 
between  the pairs $\Delta^{++,}$, $\Delta^-$ and  $\Delta^{+}$, 
$\Delta^{0}$ normalized by the mean value 
of their mass in Fig.~\ref{fig:Dm} for  $\beta=3.8$, 
$\beta=3.9$ and $\beta=4.05$. 
As can be seen, the splitting is consistent with
zero for these values of $\beta$, indicating that isospin breaking in the $\Delta$ system is small.

\section{Chiral extrapolation}
\subsection{Nucleon mass}

\begin{figure}[h]
\begin{minipage}{8cm}
\epsfxsize=8truecm
\epsfysize=6truecm
 \mbox{\epsfbox{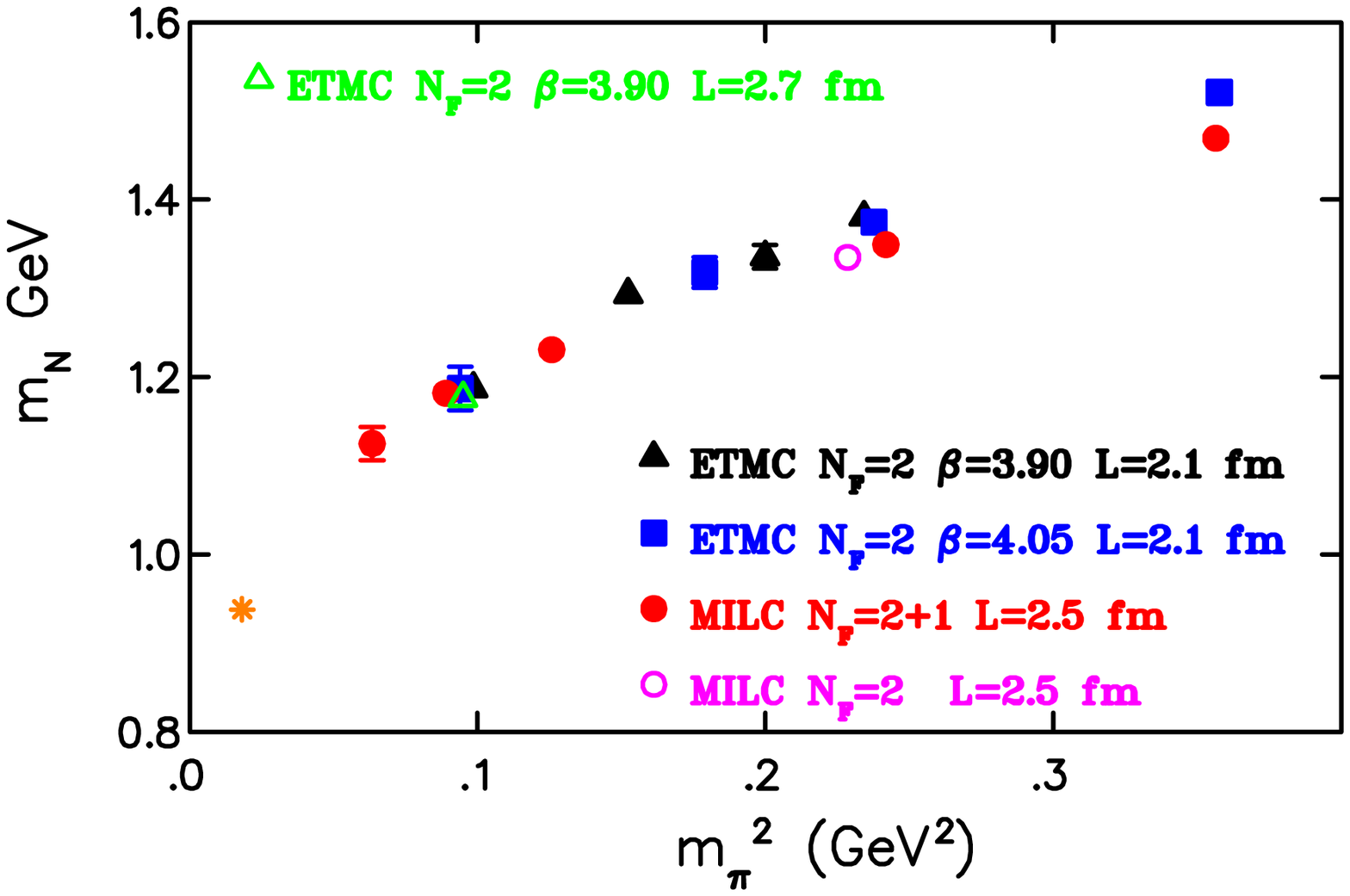}}
\caption{The nucleon mass as a function of $m_\pi^2$
for $\beta=3.9$ on a lattice of size $24^3\times 48$ (filled triangles)
and on a lattice of size $32^3\times 64$ (open triangle). Results
at $\beta=4.05$ are shown with the filled squares. The physical nucleon
mass is shown with the asterisk.
Results with dynamical staggered
fermions for $N_F=2+1$ (filled circles) and $N_F=2$ (open circle)
on a lattice of size $20^3\times 64$ with $a=0.125$~fm 
are from Refs.~\cite{staggered:1,staggered:2}.} 
\label{fig:mn}
\end{minipage}
\hfill
\begin{minipage}{8cm}
\epsfxsize=8truecm
\epsfysize=6truecm
 \mbox{\epsfbox{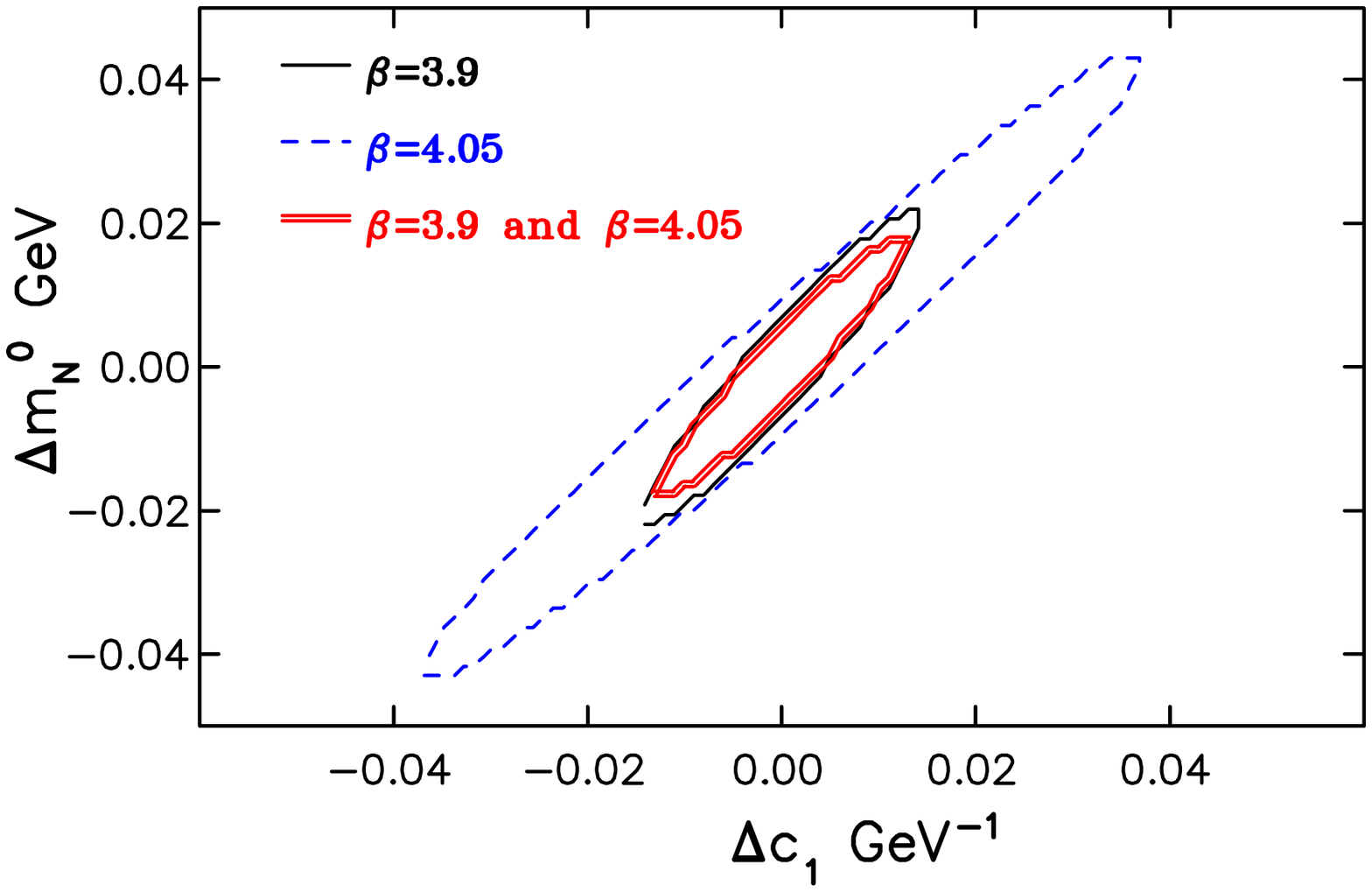}}
\caption{The variation of the fit parameters $m_N^0$ and $c_1$. The elliptical boundary
is determined by changing these parameters so that the minimal value of $\chi^2$ changes
by one. The most elongated ellipse is for $\beta=4.05$ using the nucleon mass at the three
lighter pion masses, the intermediate is for $\beta=3.9$ using all five points
and the smallest is for a combined fit to both $\beta$-values 
using a total of eight points.}
\label{fig:errorband}
\end{minipage}
\end{figure}

\begin{figure}[h]
\begin{minipage}{8cm}
\epsfxsize=8truecm
\epsfysize=10truecm
 \mbox{\epsfbox{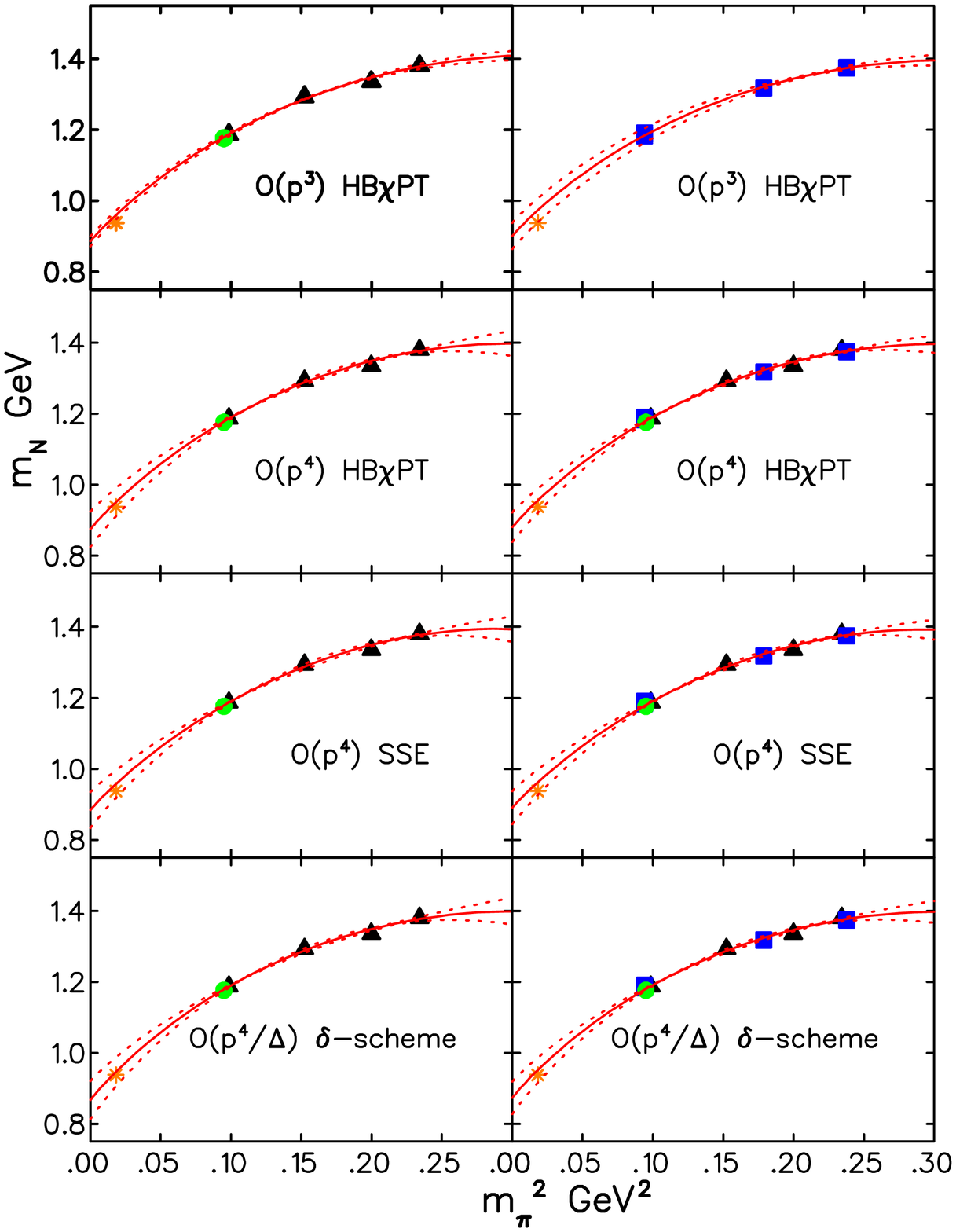}}
\caption{On the left set of graphs we show chiral fits 
to the nucleon mass for $\beta=3.9$  
using $a_{\beta=3.9}=0.0855$~fm to convert to physical units.
On the right set of graphs we show the corresponding  chiral fits
for $\beta=4.05$ using $a_{\beta=4.05}=0.0667$~fm. The upper most graph
shows the fit to  ${\cal O}(p^3)$ HB$\chi$PT where we use
our results at the three lowest
values of the pion mass. 
For the higher order fits we perform a simultaneous fit 
to both $\beta=3.9$ and $\beta=4.05$ always excluding at $\beta=4.05$ 
the result at the largest pion mass. 
 The physical point,
shown by the asterisk, is not included in the fits.}
\label{fig:chiral fit mn}
\end{minipage}
\hfill
\begin{minipage}{8cm}
\epsfxsize=8truecm
\epsfysize=10truecm
 \mbox{\epsfbox{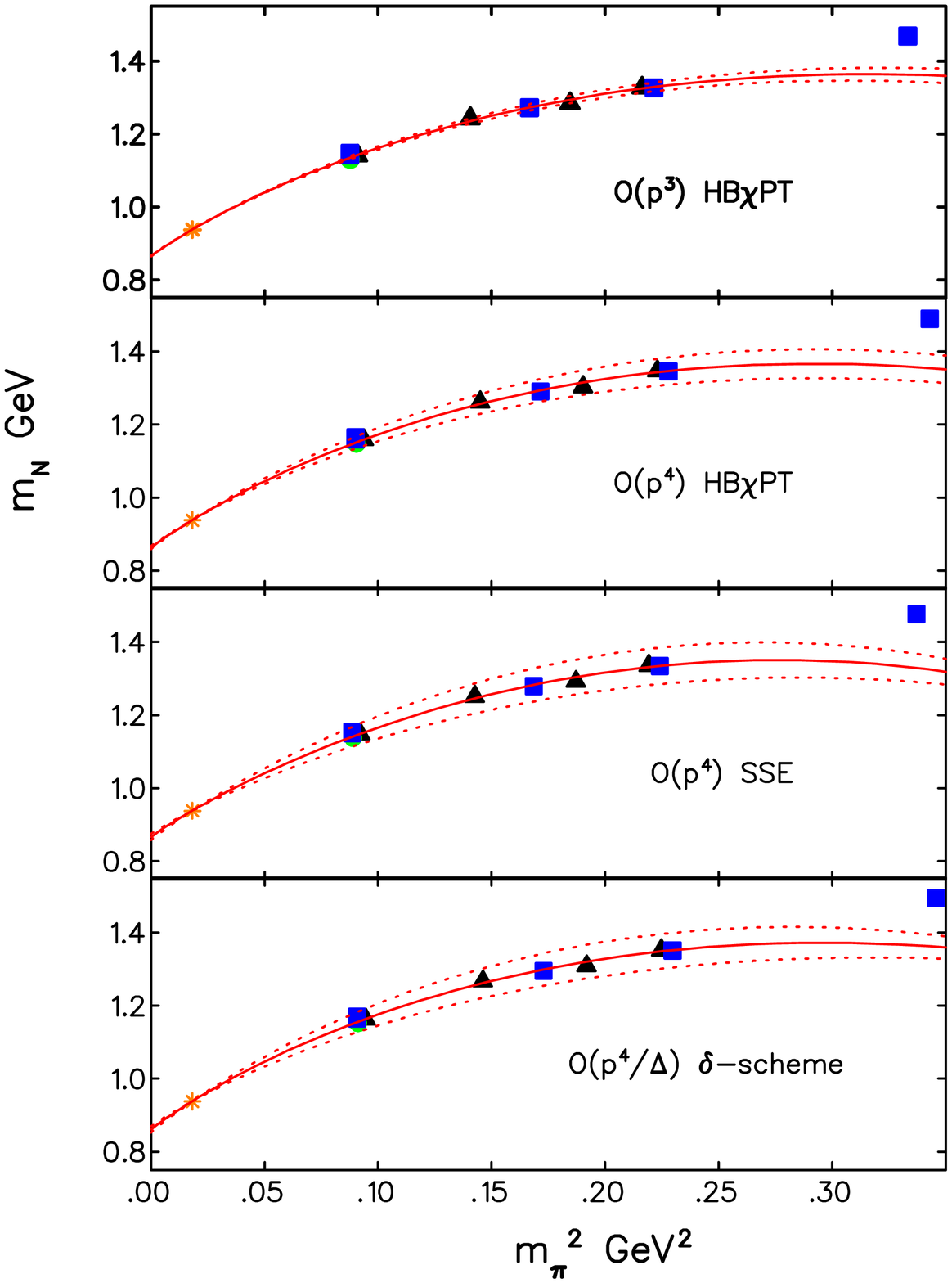}}
\caption{ Simultaneous chiral fits to the nucleon mass using results 
at $\beta=3.9$ and $\beta=4.05$ excluding the
heaviest pion value. The fits are done so that
the physical point 
shown by the asterisk is  reproduced thereby fixing the lattice spacings.
The rest of the notation is the same as that of Fig.~\ref{fig:chiral fit mn}.}
\label{fig:a from mn}
\end{minipage}
\end{figure}

\begin{figure}[h]
\begin{minipage}{8cm}
\epsfxsize=8truecm
\epsfysize=10truecm
 \mbox{\epsfbox{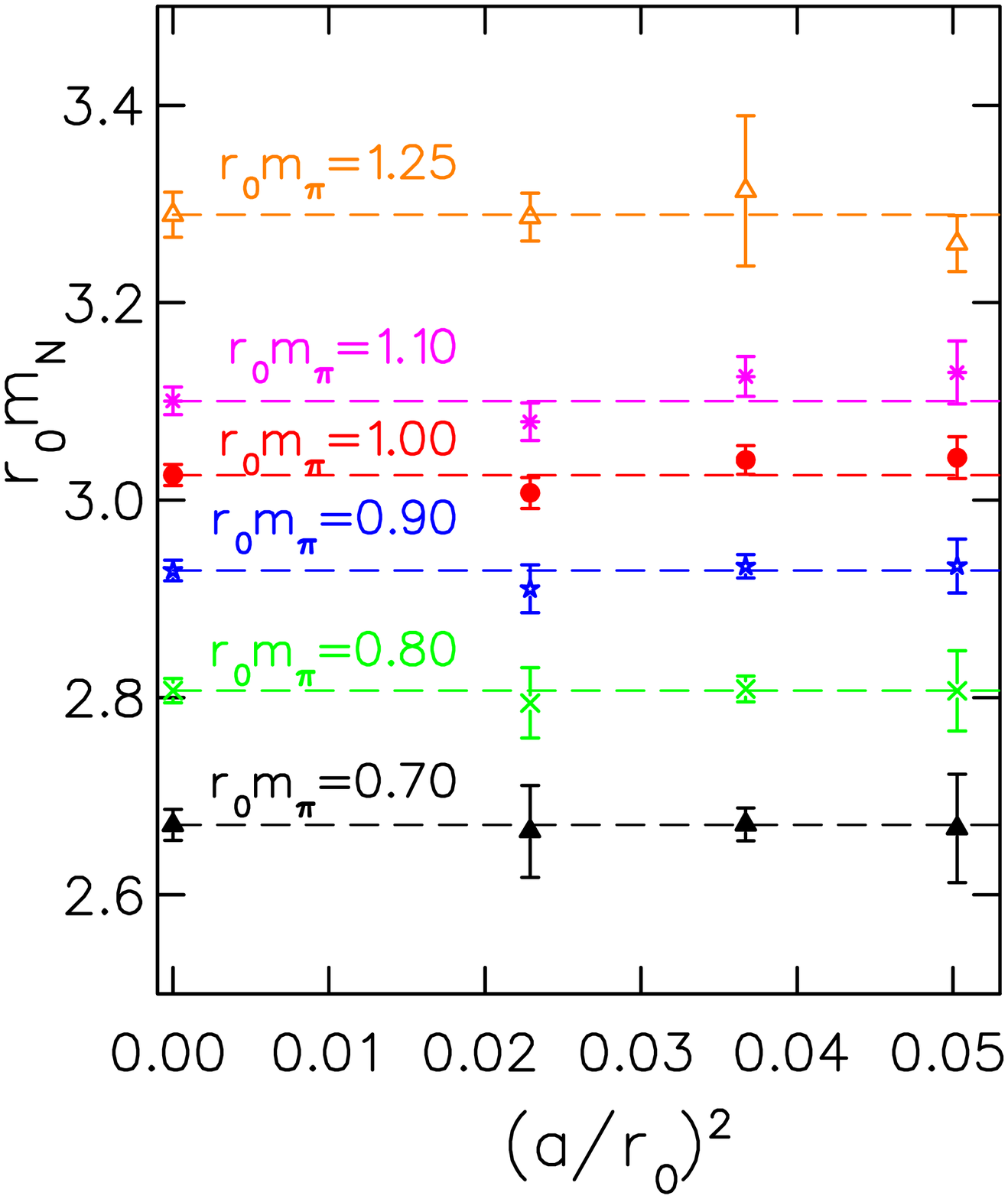}}
\caption{Continuum limit of the nucleon mass using
the lowest order HB$\chi$PT to interpolate except for the heaviest pion 
mass where we used  linear interpolation. 
For $r_0$ and $a$ we use the values determined in
the pion sector, namely $r_0/a=4.46(3),\; (r_0=0.444(4)$~fm), 
$r_0/a=5.22(2),\; (r_0=0.446(3)$~fm) and $r_0/a=6.61(3),\; (r_0=0.441(4)$~fm)
at $\beta=3.8$, $\beta=3.9$ and $\beta=4.05$ respectively~\cite{Dimopoulos:2007qy}.}
\label{fig:cont. mn}
\end{minipage}
\hfill
\begin{minipage}{8cm}
\epsfxsize=8truecm
\epsfysize=10truecm
 \mbox{\epsfbox{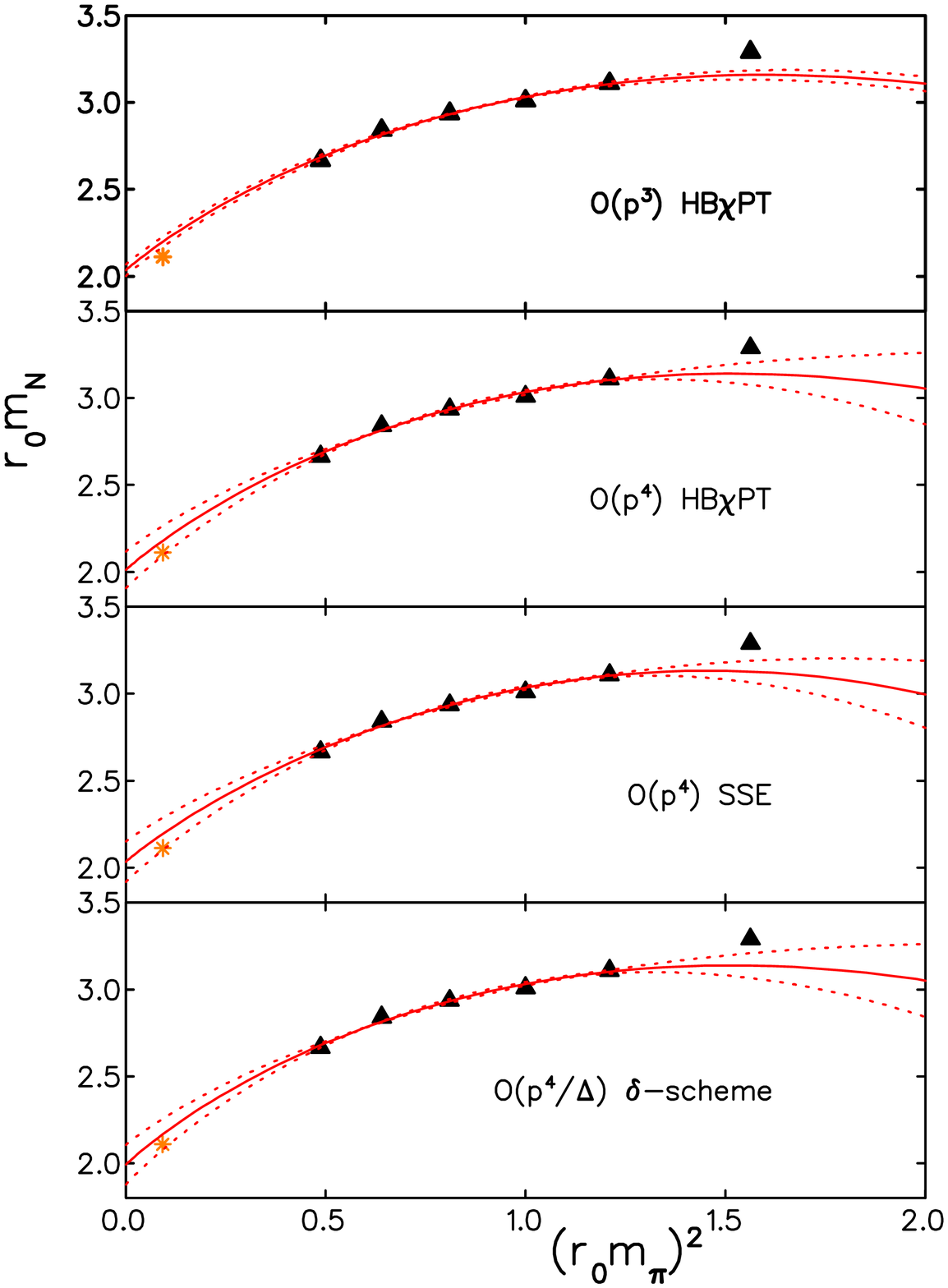}}
\caption{Chiral fits to the nucleon mass after extrapolating 
to the continuum limit 
at fixed $r_0m_\pi$ using
linear interpolation i.e. the
data given in the  second column of Table~\ref{table:interpolation}. 
The fits were done excluding the heaviest pion mass. The asterisk shows the
physical point using the estimated continuum value of $r_0=0.444$~fm. }
\label{fig:chiral fit cont. mn}
\end{minipage}
\end{figure}

We show our results for the nucleon mass as 
a function of  $m_\pi^2$
in Fig.~\ref{fig:mn}. The masses are extracted in lattice units. To convert
to physical units we need to know the value of the lattice spacing. 
A standard procedure is  to match the experimental value of  $f_\pi$ to the one obtained on the lattice extrapolated
to the physical pion mass. This gives $a=0.0855$~fm at
$\beta=3.9$ and $a=0.0667$~fm at $\beta=4.05$~\cite{Urbach:2007}. We use these values  to convert lattice results
to physical units. 
The results at these two $\beta$-values
fall on a common curve indicating that cut-off effects are small for these
values of the lattice spacing. 
In Fig.~\ref{fig:mn} we include, for comparison,
 results obtained with dynamical 
staggered fermions from Ref.~\cite{staggered:1,staggered:2}.
Results using  these two formulations  
are consistent with each other.
 As we already discussed,  results obtained on lattices of spatial length
$L_s=2.1$~fm and $L_s=2.7$~fm at $\beta=3.9$ 
for the lowest pion mass are consistent indicating
that finite volume effects are small for the pion masses used in this work.    
Therefore as a first analysis of our lattice results, we  use continuum 
chiral perturbation theory 
in an infinite volume to perform the chiral extrapolation to the
physical point. An analysis carried out after taking the continuum limit
will serve as a check of cut-off effects. 
The  leading one-loop result in heavy baryon chiral perturbation theory
(HB$\chi$PT)~\cite{Gas88} is well known:
\be 
m_N = m_N^0-4 c_1m_\pi^2- \frac{3g_A^2}{32\pi f_\pi^2} m_\pi^3
\label{mn to p3}
\ee
with $m_N^0$, the nucleon mass in the chiral limit, and $c_1$ treated as
fit parameters.
This lowest order result is the same in HB$\chi$PT with dimensional and infra-red regularization
as well as when the $\Delta$ degree of freedom is explicitly included.
It is also the same in manifestly Lorenz-invariant formulation with
infrared regularization. Therefore we will use this well established result
to predict the nucleon mass at the physical point as well as fix the
lattice spacing using the experimental nucleon mass as input. 
The higher order results will only be used to estimate the
systematic error associated with the chiral extrapolation.
We take for $f_\pi$ and $g_A$  their physical values, 
namely $f_\pi=0.092419(7)(25)$~GeV
and $g_A=1.2695(29)$, which is  what is customarily done in chiral fits to 
lattice data on the nucleon mass~\cite{Bernard:2005,Pascalutsa:2006,Procura:2006}. 
We will take the experimental values for $f_\pi$ and $g_A$ also
when using higher order results. 
In higher orders new low energy constants enter,
and we also fix their values from experimental data.
In order to determine the errors on the fit parameters we allow for 
a variation in the parameters that increases the minimal value of 
$\chi^2$ by one.
  In Fig.~\ref{fig:errorband} we show the boundary of the allowed
variation of the parameters.
 As expected when the number of available 
lattice results increases the error decreases.
In Fig.~\ref{fig:chiral fit mn} we show fits to the ${\cal O}(p^3)$ result
with the error band determined by  the maximum allowed variation in the parameters
 that increase the minimal $\chi^2$ by one. 
As can be seen, this ${\cal O}(p^3)$ result
provides a very good fit to our lattice data both at $\beta=3.9$ and $\beta=4.05$.
Since finite volume effects are small we use in the fit data on both volumes at $\beta=3.9$.
In Table~\ref{table:fit parameters}
we give the values of the  parameters 
 $m_N^0$ and $c_1$.  In this 
determination we use the lattice spacing determined from $f_\pi$. In the case of $\beta=3.9$ 
we include the result obtained using the larger volume. 
 We would like to stress that, despite the fact that the physical point is  
not included in the fit 
as customary done in other chiral extrapolations of
lattice data. The value of the nucleon mass that we find at the physical pion mass using data at both
$\beta=3.9$ and $\beta=4.05$ to fit to the ${\cal O}(p^3)$ HB$\chi$PT  of \eq{mn to p3} is 963(12)~MeV,
where the error is only statistical. 

\begin{table}[h]
  
  \begin{tabular}{lccccccccccc}
    \hline\hline
  & \multicolumn{2}{c}{$\beta=3.9$} &\multicolumn{2}{c}{  $\beta=4.05$}& \multicolumn{2}{c}{$\beta=3.9$ and $\beta=4.05$} & \multicolumn{2}{c}{Continuum}&
 \multicolumn{2}{c}{Continuum}\\
  & \multicolumn{6}{c}{} &\multicolumn{2}{c}{Linear}  & \multicolumn{2}{c}{with ${\cal O}(p^3)$ HB$\chi$PT} \\
    \hline\hline
  & $m_N^0$  & $c_1$  &  $m_N^0$  & $c_1$  & $m_N^0$  & $c_1$  & $m_N^0$  & $c_1$  & $m_N^0$  & $c_1$ \\
\hline\hline
  \multicolumn{11}{c}{ Nucleon}\\ \hline
${\cal O}(p^3)$ HB$\chi$PT           & 0.886(14) & -1.21(2)  &  0.901(37) & -1.18(4) &  0.889(13)  & -1.20(2)  &  0.904(14) & -1.19(2) & 0.898(9)  & -1.19(1) \\
${\cal O}(p^4)$ HB$\chi$PT           & 0.875(50) & -1.23(17) &  0.929     & -1.10    &  0.881(42)  & -1.22(12) &  0.893(47) & -1.21(12)& 0.889(25) & -1.21(7)\\
${\cal O}(p^4)$ SSE                  & 0.884(51) & -1.19(14) &  0.944     & -1.02    &  0.891(47)  & -1.17(15) &  0.903(52) & -1.16(15)& 0.901(30) & -1.15(9)\\
${\cal O}(p^4/\Delta) \delta$-scheme & 0.867(54) & -1.29(18) &  0.927     & -1.13    &  0.873(46 ) & -1.28(15) &  0.886(51) & -1.27(16)& 0.883(29) & -1.26(9)\\
\hline\hline
  & $m_\Delta^0$  & $c_1$  &  $m_\Delta^0$  & $c_1$  & $m_\Delta^0$  & $c_1$  & $m_\Delta^0$  & $c_1$  & $m_\Delta^0$  & $c_1$ \\
\hline\hline 
\multicolumn{11}{c}{  $\Delta^{++,-}$}\\ \hline
    \hline
 ${\cal O}(p^3)$ HB$\chi$PT          &  1.248(31) & -1.19(4) & 1.222(68)*& -1.20(5)*& 1.241(27)  & -1.21(4)  & 1.274(33) & -1.17(4)  & 1.251(16) & -1.20(2) \\
${\cal O}(p^4/\Delta) \delta$-scheme &  1.258(126)& -1.15(43)&           &          & 1.347(90)  & -0.85(30) & 1.267(80) & -1.16(20) & 1.261(54)& -1.16(17)\\ 
    \hline\hline
\multicolumn{11}{c}{  $\Delta^{+,0}$}\\ \hline
    \hline
 ${\cal O}(p^3)$ HB$\chi$PT          & 1.255(40) & -1.20(5)  & 1.261(6)* & -1.19(8)* & 1.256(33)  & -1.20(4 ) &1.264(32) & -1.18(4) & 1.262(19) & -1.19(3) \\
${\cal O}(p^4/\Delta) \delta$-scheme & 1.302(43)* & -1.03(7)* &          &           & 1.372(104) & -0.81(32) &1.373(65) & -0.85(17)& 1.267(42)& -1.16(12)\\
 \hline\hline
  \end{tabular}
  \caption{Fit parameters $m_N^0$ and $m_\Delta^0$ in GeV
 and $c_1$ in GeV$^{-1}$.
 Results with an asterisk have $\chi^2$/d.o.f. larger than one.
All fits to the continuum results excluded the largest value of $r_0 m_\pi$ with the exception of the $\Delta^{+,0}$ in the $\delta$-scheme where to obtain
a good fit we use all six points. For $\beta=3.9$ we use all masses including the 
results at the larger volume whereas for $\beta=4.05$ we use results
at the three smaller pion masses. For the fits to continuum results we give  two sets of results: the first set is obtained when using linear interpolation to the reference pion
masses and the second using ${\cal O}(p^3)$ HB$\chi$PT for interpolation.}
  \label{table:fit parameters}
\end{table}

\begin{table}[h]
  
  \begin{tabular}{lccccccccc}
    \hline\hline
&  \multicolumn{2}{c}{$\beta=3.9$} &\multicolumn{2}{c}{  $\beta=4.05$}& \multicolumn{3}{c}{$\beta=3.9$ and $\beta=4.05$} & \multicolumn{2}{c}{Continuum}\\
&  $m_N^0$  & $a_{\beta=3.9}$  & $m_N^0$  & $a_{\beta=4.05}$  & $m_N^0$  & $a_{\beta=3.9}$  & $a_{\beta=4.05}$ & $m_N^0$  & $r_0$   \\
    \hline\hline
 ${\cal O}(p^3)$ HB$\chi$PT          &  0.865(2)  & 0.0886(18) & 0.868(5) & 0.0708(37) &  0.866(1) & 0.0889(12) & 0.0691(10) & 0.868(2)  & 0.473(9) \\
${\cal O}(p^4)$ HB$\chi$PT           &  0.862(9)  & 0.0869(46) & 0.871    & 0.0717     &  0.863(4),& 0.0875(26) & 0.0681(24) & 0.863(8)  & 0.461(23) \\
${\cal O}(p^4)$ SSE                  &  0.865(10) & 0.0876(47) & 0.876    & 0.0724     &  0.866(9) & 0.0884(40) & 0.0687(31) & 0.866(9)  & 0.464(24) \\
${\cal O}(p^4/\Delta) \delta$-scheme &  0.859(11) & 0.0865(65) & 0.870    & 0.0717     &  0.861(9) & 0.0873(39) & 0.0678(31) & 0.861(10) & 0.458(25) \\
\hline\hline
  \end{tabular}
  \caption{Determination of the lattice spacing in fm and $m_N^0$ in GeV
using the nucleon mass.
 Fitting to the nucleon continuum results obtained by
linear interpolation  at the five lighter
reference pion masses we extract the continuum value of $r_0$ in fm
 by constraining the fits to reproduce the physical nucleon mass.}
  \label{table:fit a}
\end{table}

\begin{table}[h]
  
  \begin{tabular}{lccccccc}
    \hline\hline
$r_0 m_\pi$&  \multicolumn{2}{c}{$r_0 m_N$} &\multicolumn{2}{c}{  $r_0 m_{\Delta^{++,-}}$}& \multicolumn{2}{c}{$r_0 m_{\Delta^{+,0}}$}\\
    \hline\hline
\multicolumn{7}{c}{$\beta=3.8$}\\
\hline\hline
 0.70 & 2.654(95) & 2.668(55) &   3.596(119)&  3.511(69) &  3.502(161) &  3.528(79) \\
 0.80 & 2.804(45) & 2.807(41) &   3.637(54) &  3.614(47) &  3.641(65)  &  3.652(53) \\
 0.90 & 2.935(40) & 2.933(27) &   3.667(75) &  3.701(28) &  3.790(111) &  3.760(30) \\
 1.00 & 3.044(45) & 3.043(21) &   3.731(83) &  3.767(31) &  3.885(120) &  3.850(33) \\
 1.10 & 3.133(34) & 3.129(31) &   3.831(63) &  3.804(57) &  3.918(72)  &  3.915(64) \\
 1.25 & 3.256(28) & 3.201(63) &   4.086(62) &  3.793(111)&  4.127(64)  &  3.950(127) \\
\hline\hline
\multicolumn{7}{c}{$\beta=3.9$}\\  
 0.70 & 2.666(27) & 2.672(17)  & 3.548(60)  & 3.481(35)  &  3.632(88)  & 3.498(46) \\
 0.80 & 2.840(27) & 2.809(13)  & 3.621(65)  & 3.614(25)  &  3.641(61)  & 3.633(32) \\
 0.90 & 2.947(22) & 2.933(12)  & 3.720(56)  & 3.734(16)  &  3.725(52)  & 3.754(20) \\
 1.00 & 3.013(31) & 3.041(14)  & 3.841(60)  & 3.837(18)  &  3.861(46)  & 3.859(17) \\
 1.10 & 3.132(24) & 3.125(20)  & 3.917(50)  & 3.916(30)  &  3.943(50)  & 3.938(31) \\
 1.25 & 3.313(76) & 3.194(33)  & 4.028(66)  & 3.976(56)  &  4.061(105) & 4.000(62) \\
 \hline\hline
\multicolumn{7}{c}{$\beta=4.05$}\\  
 0.70 & 2.676(56) & 2.665(47)  & 3.565(124) &  3.428(81) &  3.550(97) & 3.477(71) \\
 0.80 & 2.843(71) & 2.795(36)  & 3.627(137) &  3.571(59) &  3.627(107) & 3.609(51) \\
 0.90 & 2.903(40) & 2.910(24)  & 3.669(45)  &  3.701(35) &  3.690(45) & 3.727(30) \\
 1.00 & 3.003(40) & 3.007(16)  & 3.771(45)  &  3.815(18) &  3.793(46) & 3.827(18) \\
 1.10 & 3.086(24) & 3.079(19)  & 3.932(43)  &  3.905(35) &  3.936(50) & 3.902(37) \\
 1.25 & 3.287(24) & 3.126(42)  & 4.017(48)  &  3.982(85) &  4.038(49) & 3.953(84) \\
 \hline\hline
\multicolumn{7}{c}{Continuum}\\  
 0.70  &  2.667(24) &  2.671(16) & 3.551(54) & 3.472(32)  &  3.595(65) &  3.492(38) \\
 0.80  &  2.840(25) &  2.807(12) & 3.622(59) & 3.608(23)  &  3.638(53) &  3.626(27) \\
 0.90  &  2.936(19) &  2.929(11) & 3.687(35) & 3.728(15)  &  3.705(34) &  3.746(16) \\
 1.00  &  3.009(25) &  3.025(11) & 3.794(36) & 3.826(13)  &  3.827(33) &  3.844(12) \\
 1.10  &  3.109(17) &  3.101(14) & 3.926(33) & 3.916(23)  &  3.939(35) &  3.924(24) \\
 1.25  &  3.289(23) &  3.168(26) & 4.021(39) & 3.978(47)  &  4.042(44) &  3.984(50) \\
 \hline\hline
  \end{tabular}
  \caption{Results for the nucleon,  $\Delta^{++,-}$ and $\Delta^{+,0}$
mass interpolated at the same value of $r_0 m_\pi$ for the three $\beta$ values.
The continuum limit is then taken at constant $r_0 m_\pi$ using the results
at $\beta=3.9$ and $\beta=4.05$. We give results using linear interpolation in 
second, fourth and sixth columns whereas in the third, fifth and seventh columns we give
the results using  
lowest order HB$\chi$PT.}
  \label{table:interpolation}
\end{table}

The nucleon sigma term is defined by
\be
\sigma_N=\sum_{q=u,d} \mu_q \frac{dM_N}{d\mu_q}
\label{sigma term}
\ee
where we have neglected contributions from other quarks. Following
Ref.~\cite{Procura:2003ig} we use the relation
$m_\pi^2\sim \mu$ to evaluate $\sigma_N$ by computing 
$m_\pi^2\frac{dM_N}{dm_\pi^2}$. Using the value of $c_1$ determined from
the nucleon fit we find at the physical point $\sigma_N=66.7\pm 1.3$~MeV, where
the error  is statistical. This value 
is larger than the prevailing value of $45\pm 8$~MeV~\cite{Gas91} 
but in agreement
with a new analysis~\cite{Pav02} that includes additional 
data.
Our current calculation does not include a dynamical strange quark
and a better understanding of this term could come
when simulations with dynamical strange
quarks are available~\cite{Gas82}. 
Note that given the role of the sigma term for what
concerns the chiral extrapolation as well as its implication in dark matter
detection~\cite{Ell08} it is clear that a serious effort to better fix its
experimental value is highly desirable.

Chiral corrections to the nucleon mass are known to 
 ${\cal O}(p^4)$ within several
expansion schemes.
In HB$\chi$PT to  ${\cal O}(p^4)$ 
 with dimensional 
regularization~\cite{Steininger98,Bernard:2004,Bernard:2005} the result is given by 
\beq
m_N & = & m_N^0-4 c_1m_\pi^2- \frac{3g_A^2}{32\pi f_\pi^2} m_\pi^3 -4 E_1(\lambda)m_\pi^4 
        +\frac{3 m_\pi^4}{32 \pi^2 f_\pi^2}\biggl[\frac{1}{4} \left (c_2-\frac{2g_A^2}{m_N^0}\right)
- \left(c_2-8c_1+4c_3+\frac{g_A^2}{m_N^0}\right) \log\left (\frac{m_\pi}{\lambda}\right)\biggr]\quad .
\label{HBchi2}
\eea
We take the cut-off scale $\lambda=1$~GeV and fix the dimension two low energy  constants 
$ c_2=3.2$~GeV$^{-1}$~\cite{Fettes:1998} and $c_3=-3.45$~GeV$^{-1}$~\cite{Bernard:2004,Procura:2006}.  This value is consistent with empirical
nucleon-nucleon phase shifts~\cite{Entem:2002sf,Epelbaum:2004}.
The counter-term $E_1$ is taken as an additional fit parameter.
 HB$\chi$PT with dimensional 
regularization is in agreement with
covariant baryon $\chi$PT with infrared regularization
up to a recoil term given by 
$\frac{3g_A^2 m_\pi^5}{256 \pi f_\pi^2 {m_N^0}^2} $
that is of no numerical significance~\cite{Procura:2006}. We have included this
term in our fits.  
In the so called small scale expansion (SSE)~\cite{Procura:2006},
the
$\Delta$-degrees of freedom are explicitly included in covariant baryon 
$\chi$PT 
by introducing as an additional counting parameter the
$\Delta$-nucleon mass
splitting, $\Delta\equiv m_\Delta-m_N$,  taking
${\cal O}(\Delta/m_N)\sim {\cal O}(m_\pi/m_N)$. In  SSE the nucleon mass is given by
\bea
        m_N &= &m_N^0 - 4c_1 m_\pi^2 - \frac{3g_A^2}{32\pi f_\pi^2} m_\pi^3 -4 E_1(\lambda)m_\pi^4 
            -\frac{3\left(g_A^2+3c_A^2\right)}{64\pi^2 f_\pi^2 m_N^0} m_\pi^4
          -\frac{(3g_A^2+10c_A^2)}{32\pi^2 f_\pi^2 m_N^0} m_\pi^4\log\left (\frac{m_\pi}{\lambda}\right)\nonumber \\
           & & - \frac{c_A^2}{3\pi^2 f_\pi^2} \left(1+\frac{\Delta}{2m_N^0}\right)
                   \biggl[ \frac{\Delta}{4}m_\pi^2 
                  + \left(\Delta^3-\frac{3}{2} m_\pi^2 \Delta\right) \log\left (\frac{m_\pi}{2\Delta}\right ) 
                   + (\Delta^2-m_\pi^2)  R(m_\pi) \biggr]
\label{SSE}
\eea
where $R(m_\pi)=\sqrt{m_\pi^2-\Delta^2}\cos^{-1}\left(\frac{\Delta}{m_\pi}\right)$ if $m_\pi>\Delta$
and $R(m_\pi)=\sqrt{\Delta^2-m_\pi^2} \log\left( \frac{\Delta}{m_\pi}+\sqrt{\frac{\Delta^2}{m_\pi^2}-1} \right)$ for $m_\pi \le \Delta$.
We take $c_A=1.127$~\cite{Procura:2006}, $\lambda=1$~GeV and fit the counter-term $E_1$.
A different counting scheme, 
known as $\delta$-scheme, takes
$\Delta/m_N\sim {\cal O}(\delta)$ and
$m_\pi/m_N\sim{\cal O}(\delta^2)$~\cite{Pascalutsa:2006}.
 Using the $\delta-$scheme in a covariant chiral expansion to order
 $p^4/\Delta$
 one obtains an expansion that has a similar form 
for the nucleon and $\Delta$ mass. The nucleon mass is given by
\be
     m_N = m_N^0 - 4c_1 m_\pi^2 - \frac{1}{2} \frac{{m_N^0}^3}{(8\pi f_\pi)^2} \Biggl [
                        9g_A^2  V_{\rm loop}\left (\frac{m_\pi}{m_N^0},0\right) 
+ \frac{4 h_A^2}{\left(1+\frac{\Delta}{m_N^0}\right)^2} V_{\rm loop}\left(\frac{m_\pi}{m_N^0},\frac{\Delta}{m_N^0}\right)
 \Biggr ] + c_2  m_\pi^4
\label{delta-scheme}
\ee
The $\pi N$ and $\pi\Delta$ loop function $V_{\rm loop}$  is
 given in Ref.~\cite{Pascalutsa:2006} and, following the same 
reference, we take the value of  $h_A=2.85$.
Here we use the variant of the $\delta$-scheme that includes the 
$\pi\Delta$-loop and adds the fourth order term  $c_2 m_\pi^4$ 
as an estimate of higher order effects, since the complete fourth order
result is not available.
The parameter $c_2$ is to be determined from  the lattice data.
 The fits using these different formulations are shown in 
Fig.~\ref{fig:chiral fit mn}. At $\beta=3.9$ shown in the left panel
we used $a_{\beta=3.9}=0.0855$~fm to convert to physical units. We have
four values for the the $24^3\times 48$ lattice and one for
the larger lattice. 
At $\beta=4.05$ we only use results at the three smallest
pion masses since including the result at the largest pion mass 
yields fits with unacceptably large $\chi^2/{\rm d.o.f.}$. 
Therefore only at lowest order
$\chi$PT where we have only two fitting parameters
 we can perform a fit. For the higher order we give the values
of $m_N^0$ and $c_1$ of the curves that pass through all the lattice points.
Since cut-off effects are consistent with zero for these two values of $\beta$
we can use these two sets of results in a combined fit. 
For the lattice data at $\beta=4.05$ we use $a_{\beta=4.0}=0.0667$~fm
determined from $f_\pi$, to convert to physical units.
The experimental
value of the nucleon is shown with the asterisk.
In Table~\ref{table:fit parameters} we  give the values 
of $m_N^0$ and $c_1$ when simultaneous fits to both
$\beta=3.9$ and $\beta=4.05$ data are done.  We use the lattice spacings
determined from the pion decay constant to convert to physical 
units~\cite{Urbach:2007}.
These fits are shown in Fig.~\ref{fig:chiral fit mn} when
using higher order $\chi$PT.
All formulations provide a good description of the lattice results
and yield a nucleon mass at the physical point consistent with  
the experimental value. As already discussed, the  value of the nucleon mass that we find using \eq{mn to p3}  is 963(12)~MeV.
The corresponding value using ${\cal O}(p^4)$ HB$\chi$PT is 955(33) MeV. We take the difference between these two mean values
as an estimate of the systematic error due to the chiral extrapolation and quote $963\pm 12 (stat.) \pm 8 (syst.)$~MeV
as our prediction of the nucleon mass. Within the statistical and estimated systematical uncertainty this value is close
to the experimental one. 
Furthermore the values extracted for the nucleon at 
the chiral limit $m_N^0$ as well as $c_1$
are in agreement in all formulations.
In addition the value of  nucleon $\sigma_N$ term defined in \eq{sigma term} 
can be evaluated using HB$\chi$PT to ${\cal O}(p^4)$ of \eq{HBchi2}. 
If we use the next to leading order
 relation between
$m_{\pi}^2$ and the quark mass
 $\mu_q$~\cite{Foster:1998vw, Boucaud:2006,Urbach:2007} instead of
the leading order relation $ m_{\pi}^2 \propto \mu_q$ we find a value of
   $ 67 \pm 8.0$ MeV
at the physical point, which is consistent with the value
obtained to ${\cal O}(p^3)$ albeit with a larger error. We note that
the impact on $\sigma_N$ of using \eq{sigma term}
 with the next to leading order result,
rather than the lowest oder relation, between $m_{\pi}^2$ and $\mu_q$ is
small and yields a relative decrease of its value at the physical
point of about 2\% only.

The consistency between the ${\cal O}(p^3)$ result and the higher order  expansions
allows for an
extrapolation to the physical point and a determination of 
the lattice spacing using  the nucleon mass. Fixing the lattice spacing
from the nucleon mass allows for
a comparison with the value obtained from the pion sector and provides a non-trivial
check of our lattice formulation. 
We consider $a_{\beta=3.9}$ and $a_{\beta=4.05}$ 
as independent fit parameters in a combined fit of data at $\beta=3.9$ and 
 $\beta=4.05$ where the physical nucleon mass is included with no error.
In this way the lattice spacings can be determined solely by using as input 
the nucleon mass at the physical point.
Using the leading one-loop result  
 we find  
$a_{\beta=3.9}=0.0889(12)$~fm and $a_{\beta=4.05}=0.0691(10)$~fm.
The quality of these fits are shown in  Fig.~\ref{fig:a from mn}.
The values of the lattice spacing  obtained 
to ${\cal O}(p^4)$ using Eq.~(\ref{HBchi2})
 are given in Table~\ref{table:fit a}. Both SSE
and  the $\delta-$scheme defined by  Eqs.~(\ref{SSE})
and (\ref{delta-scheme}),  which
include explicitly $\Delta$-degrees of  freedom,
 yield values that are  consistent with those obtained
in  HB$\chi$PT.
The variation in the value of $a$ in the
different chiral extrapolation schemes gives an estimate
of the systematic error involved in the chiral extrapolation.
We take the difference between the mean values obtained using ${\cal O}(p^3)$
and  ${\cal O}(p^4)$ HB$\chi$PT as an estimate of the  
systematic error. Our lattice spacings
fixed using the nucleon mass are 
therefore $a_{\beta=3.9}=0.0889\pm 0.0012(stat.)\pm 0.0014 (syst.)$~fm 
and $a_{\beta=4.05}=0.0691\pm 0.0010(stat.)\pm 0.0010 (syst.)$~fm and are
on the upper bound of the values obtained using $f_\pi$.

The physical spatial volumes of the $24^3$ lattice at $\beta=3.9$  and that of the $32^3$ lattice at $\beta=4.05$
 are about $(2.1)^3$~fm$^3$. Bearing in mind that volume corrections 
for this lattice size are shown to be small
we use results obtained on these two almost equal volumes 
to estimate our masses at the continuum limit. 
In order to take the continuum limit we interpolate data, expressed in units
of $r_0$, at the same value
of $r_0 m_\pi$, where the Sommer parameter  $r_0$ is determined from the force between two static
quarks.  We use  $r_0/a= 4.46(3)$,  $r_0/a= 5.22(2)$ and  $r_0/a= 6.61(3)$
for $\beta=3.8$, $\beta=3.9$ 
and $\beta=4.05$ respectively~\cite{Dimopoulos:2007qy}.
 The values of $r_0m_\pi$ that we choose are close to  the pion mass values
where our computation is done and 
are  given in Table~\ref{table:interpolation}. We use a linear interpolation
or the fit curves determined using chiral effective theories to obtain the nucleon
mass at these reference values of $r_0 m_\pi$. 
This procedure is
done for our three $\beta$-values. We use the results at constant $r_0 m_\pi$
at $\beta=3.9$ and $\beta=4.05$ to estimate the continuum limit by fitting
to a constant under the assumption that residual cut-off effects on the pion and nucleon masses
as well as on $r_0$ at
$\beta=3.9$ and $\beta=4.05$ are negligible. This assumption is corroborated by our lattice data shown in Fig.~\ref{fig:mn} 
Results at $\beta=3.8$ at the same value of $r_0 m_\pi$ serve as a
check for the consistency of this procedure. This is illustrated in Fig.~\ref{fig:cont. mn}
where results at $\beta=3.8$ are consistent with the constant fit. 
 The results at the
continuum limit are then chirally extrapolated. The parameters obtained are
given in Table~\ref{table:fit parameters} and the fits
are shown in Fig.~\ref{fig:chiral fit cont. mn} where we excluded the heaviest
pion mass from these fits.  
The values of $m_N^0$ and $c_1$ obtained from the fits to continuum results
are consistent with the values obtained using results at finite $a$. 
This
demonstrates that cut-off effects are small.

The value of the parameter $r_0$ can be determined from our 
results in the continuum limit using the value of the physical nucleon mass. 
We give the extracted values in Table~\ref{table:fit a} 
where we used  linear interpolation to obtain the nucleon mass
at the reference values of $r_0 m_\pi$. 
  The values extracted
for $r_0$ in the continuum limit using these fits are consistent. Had we used
chiral fits  at ${\cal O}(p^3)$ to interpolate
 the value extracted would  change by $0.004$~fm and
at ${\cal O}(p^4)$ by $0.002$~fm. 
These changes are smaller than the statistical errors.
We again take the variation in the value of $r_0$ at ${\cal O}(p^3)$
and ${\cal O}(p^4)$ HB$\chi$PT as an estimate of the systematic error due to
the chiral extrapolation. Using the values given in Table~\ref{table:fit a} 
 this difference is $0.012$~fm.
We add to this error  the variation in the
values obtained using a linear interpolation scheme and the ${\cal O}(p^3)$ fit, which
  is $0.004$~fm.
Therefore the
 value that we find is $r_0=0.473 \pm 0.09(stat.) \pm 0.016(syst.)$~fm.
This value of $r_0$, like for the lattice spacing, is at the upper bound of the
value $r_0=0.444(3)$~fm~\cite{Urbach:2007} determined using $f_\pi$. The validation of
these consistency checks 
 suggests that lattice artifacts
that can affect the value of the lattice spacing when using different 
observables are small.

\subsection{$\Delta$ mass}
We perform a similar analysis as 
for the nucleon mass in the case
of the $\Delta$. 
\begin{figure}[h]
\begin{minipage}{8cm}
\epsfxsize=8truecm
\epsfysize=7truecm
 \mbox{\epsfbox{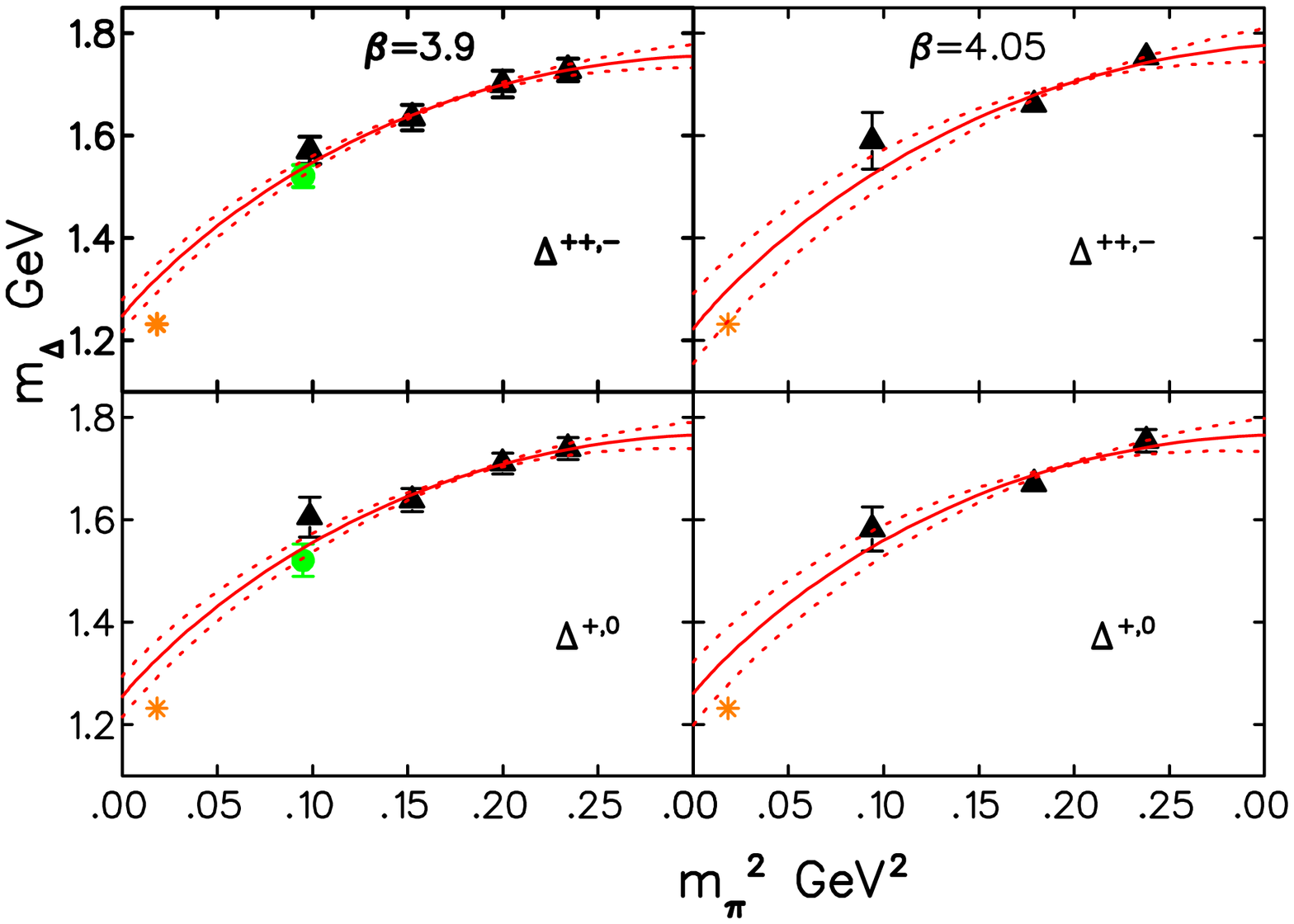}}
\caption{Chiral fits to the  $\Delta^{++,-}$ and $\Delta^{+,0}$ mass 
using Eq.~(\ref{delta fit0}) taking $a_{\beta=3.9}=0.0855$~fm
and $a_{\beta=4.05}=0.0667$~fm determined from $f_\pi$.
Filled triangles show results on a $L_s=2.1$~fm. The result on the $2.7$~fm
volume at $\beta=3.9$ is shown with the filled circle.
}
\label{fig:chiral fit delta}
\end{minipage}
\hfill
\begin{minipage}{8cm}
\epsfxsize=8truecm
\epsfysize=7truecm
 \mbox{\epsfbox{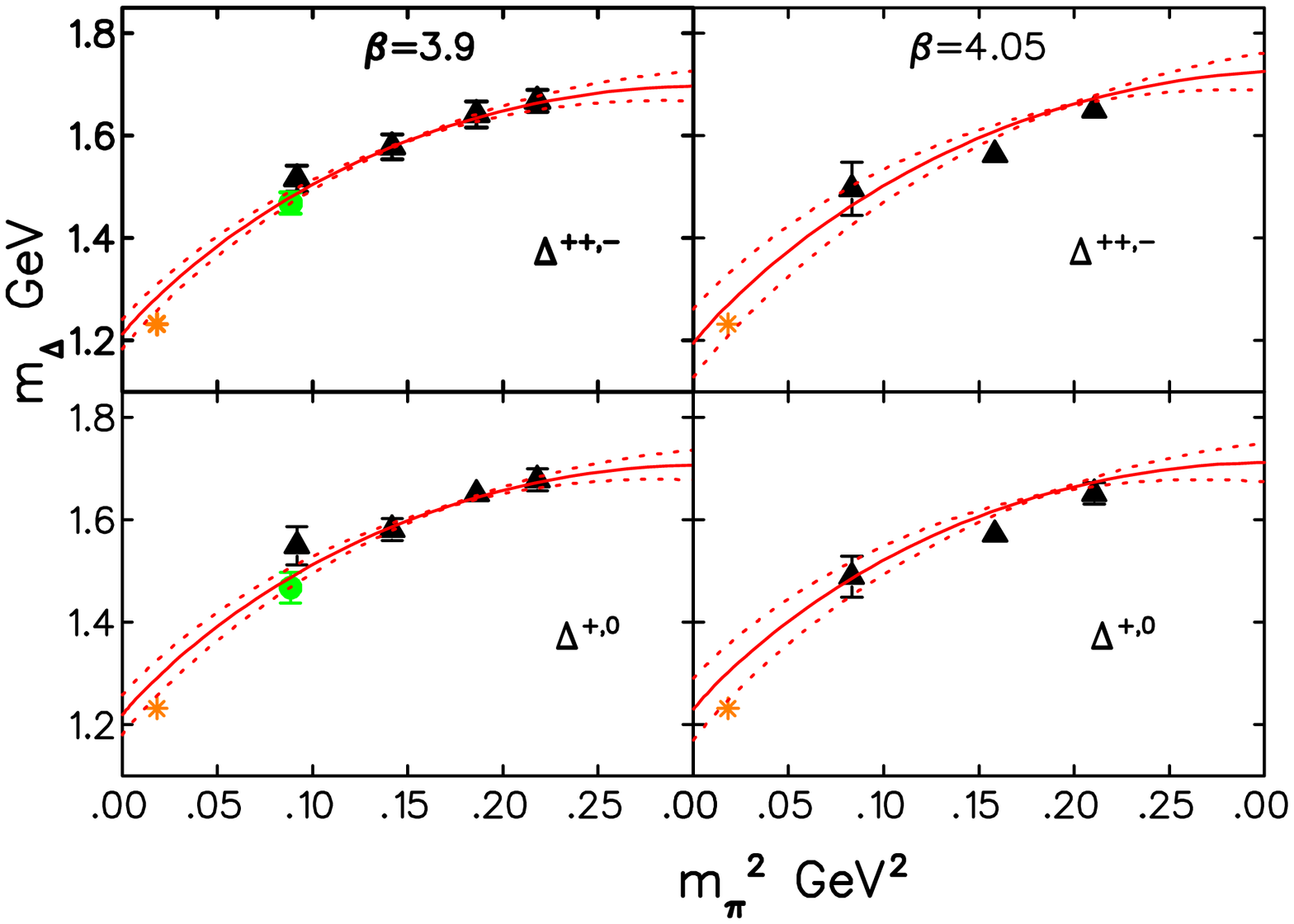}}
\caption{Chiral fits to the $\Delta^{++,-}$ and $\Delta^{+,0}$ mass
using Eq.~(\ref{delta fit0})
 with the lattice spacing determined from the nucleon mass.
 The physical point
shown by the asterisk is not included in the fits.}
\label{fig:chiral fit delta anuc}
\end{minipage}
\end{figure}

\begin{figure}[h]
\begin{minipage}{8cm}
\epsfxsize=8truecm
\epsfysize=10truecm
 \mbox{\epsfbox{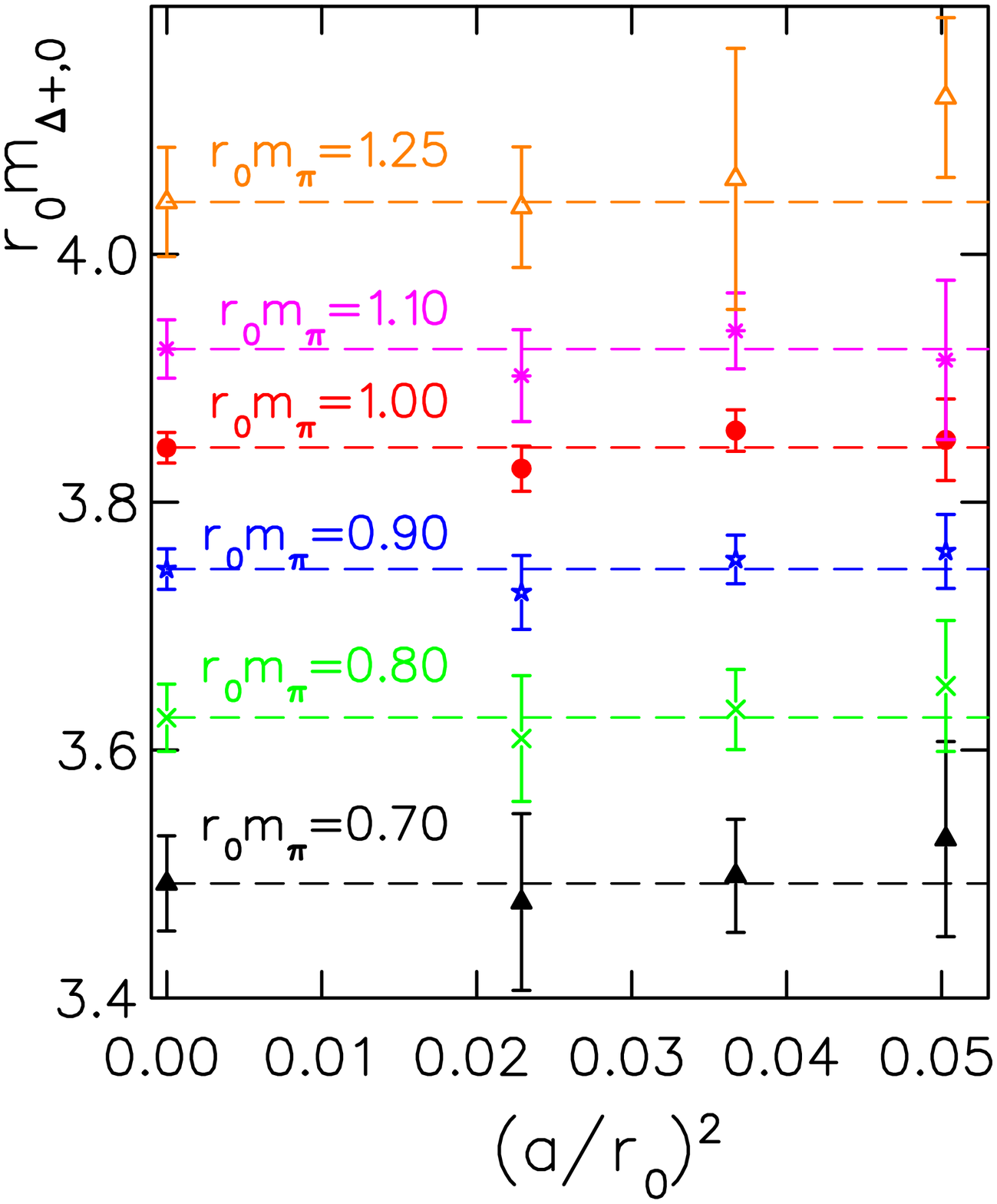}}
\caption{Continuum limit of the $\Delta^{+,0}$ mass using
the lowest order HB$\chi$PT to interpolate except for the heaviest pion 
mass where we use  linear interpolation.}
\label{fig:cont. delta}
\end{minipage}
\hfill
\begin{minipage}{8cm}
\epsfxsize=8truecm
\epsfysize=10truecm
 \mbox{\epsfbox{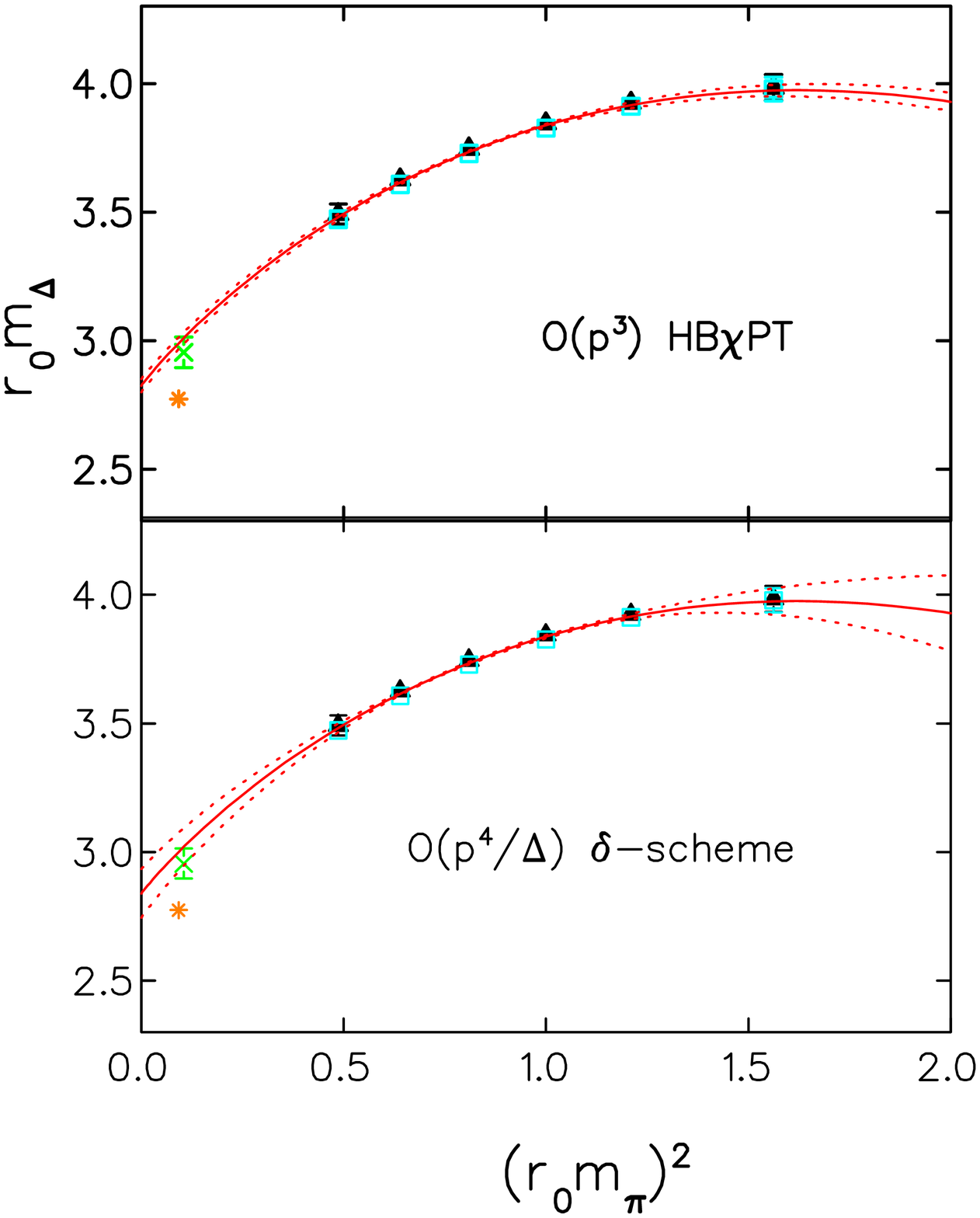}}
\caption{Simultaneous chiral fits to the $\Delta^{++,-}$ and $\Delta^{+,0}$
 mass after extrapolating to the continuum limit
at fixed $r_0 m_\pi$ excluding the
heaviest pion mass. The asterisk denotes the physical $\Delta$ mass
using the $r_0=0.444(3)$~fm extracted in the pion sector and it is not
included in the fits. The cross denotes the physical point
when we use the value of $r_0$ extracted from the nucleon sector
with only statistical errors.}
\label{fig:chiral fit cont. delta}
\end{minipage}
\end{figure}

The leading one-loop HB$\chi$PT result  in the case of the $\Delta$ mass
has the same form as that for the nucleon mass and is given by
\be
m_\Delta = m_\Delta^0-4 c_1m_\pi^2- \frac{25}{81}\frac{3H_A^2}{32\pi f_\pi^2} m_\pi^3
\label{delta fit0}
\ee
where $ m_\Delta^0$, is the $\Delta$ mass at
the chiral limit and 
$c_1 $  now denotes the coefficient 
 of the $m_\pi^2$-term for the $\Delta$ mass.  
 For the $\Delta$ axial coupling, $H_A$, 
we use the SU(6) relation $H_A=(9/5)g_A$ and 
therefore the one-loop contribution 
takes the same numerical value  as in the nucleon case.
We also consider the $\delta-$scheme to order ${\cal O}(p^4/\Delta)$
which yields an expression  that is similar to  the nucleon case:
\be
     m_\Delta = m_\Delta^0 - 4c_1 m_\pi^2 - \frac{1}{2} \frac{{m_\Delta^0}^3}{(8\pi f_\pi)^2} \Biggl [
                        g_A^2  V_{\rm loop}\left( \frac{m_\pi}{m_\Delta^0},0\right) 
                          + 4 h_A^2 V_{\rm loop}\left(\frac{m_\pi}{m_\Delta^0},\frac{\Delta}{m_\Delta^0}\right) \biggr ]
                + c_2  m_\pi^4 \quad.
\label{delta-scheme2}
\ee

The fits  using the ${\cal O}(p^3)$ HB$\chi$PT result at $\beta=3.9$ 
and $\beta=4.05$ are
shown in Fig.~\ref{fig:chiral fit delta} for the  $\Delta^{++,-}$ 
and $\Delta^{+,0}$ masses
using the lattice spacings determined from $f_\pi$. 
It is useful to chirally extrapolate the $\Delta$ mass
to see how close current results are to $\Delta(1232)$ taking
the lattice spacings as determined from the nucleon mass. 
The fits in this case are 
shown in Fig.~\ref{fig:chiral fit delta anuc} again using the ${\cal O}(p^3)$ HB$\chi$PT result.
Agreement with the experimental
value of the $\Delta$ mass
is better when one uses the lattice spacing determined from the 
nucleon mass. This indicates that for baryonic observables is favorable to
use the lattice spacing determined from the nucleon mass.
We give the values of the parameters that we extract in Table~\ref{table:fit parameters}.

The continuum extrapolation is carried out as in the nucleon case. We show 
in Fig.~\ref{fig:cont. delta} the results  for the three
different lattice spacings. As in the case of the nucleon,
the continuum limit found by averaging results from $\beta=3.9$ and $\beta=4.05$
is consistent with  results at $\beta=3.8$.
Furthermore, we find that  in the continuum limit  $\Delta^{++,-}$ and $\Delta^{+,0}$
are degenerate within errors, a result that corroborates 
absence of  isospin breaking. We therefore perform 
simultaneous fits to both $\Delta^{++,-}$ and $\Delta^{+,0}$ mass using our continuum
limit  results at the five smallest pion reference masses. 
These fits using leading chiral perturbation theory and
the $\delta-$scheme are  shown in Fig.~\ref{fig:chiral fit cont. delta}.
The physical point is again  not included in the fits. We find a $\Delta$
 mass at the physical point that is very close to experiment. Again, this agreement
improves
when we use the value of $r_0$ fixed from the nucleon mass. 
  The values that we find for $m_\Delta^0$ and $c_1$ from these simultaneous fits
using the continuum results are in good agreement with the values extracted for
$\Delta^{++,-}$ and $\Delta^{+,0}$ at finite lattice spacing. This points to
small cut-off effects and to isospin breaking effects that are smaller than statistical errors.

\section{Conclusions}
Using dynamical 
twisted mass fermions we obtain  accurate results on the nucleon
mass for pion masses in the range of  300-500 MeV.
The quality of these results 
allows a chiral extrapolation using heavy baryon
 chiral perturbation theory to
 ${\cal O}(p^3)$. There is agreement among
all approaches for this lowest order result.
Performing a simultaneous fit to our results at the two finer
lattice spacings we find a value of the nucleon mass of
 $0.963\pm 0.012 (stat.)\pm 0.008 (syst.)$~GeV where the systematic error
is the difference between the mean values  obtained at  ${\cal O}(p^3)$
and  ${\cal O}(p^4)$ HB$\chi$PT.
Comparing results at our three $\beta$-values and at the
continuum limit we confirm that
cut-off effects are small. 
 Given that this leading one-loop
result in HB$\chi$PT yields good fits to our lattice data we
use it to extract the lattice spacing from the nucleon mass at
the physical point. We find  
$a_{\beta=3.9}=0.0889\pm 0.0012(stat.)\pm 0.0014 (syst.)$~fm 
and $a_{\beta=4.05}=0.0691\pm 0.0010(stat.)\pm 0.0010 (syst.)$~fm.
Again the systematic errors are estimated by
comparing the value obtained at lowest order to the results
obtained using the ${\cal O}(p^4)$ in HB$\chi$PT.
Within this estimated uncertainty of the chiral extrapolation,
the value we find for the lattice spacings $a_{\beta=3.9}$ and 
$a_{\beta=4.05}$ is consistent with
the value  determined from $f_\pi$.
A combined analysis of data in the pion and nucleon sector
is a promising option that will be considered in the future.
We use continuum extrapolated results to determine also the
value of $r_0$ using the nucleon mass at the physical point.
We find a value of $r_0=0.473\pm 0.09 (stat.)\pm 0.016(syst.)$~fm
which is, within errors, consistent with the value determined from $f_\pi$.
The confirmation that isospin
breaking in the $\Delta$ is consistent with zero is a very important conclusion
of this work. This is demonstrated by 
evaluating the  mass splitting in the $\Delta$ isospin multiplets 
for three lattice spacings on two volumes. Consequently the mass of
the $\Delta^{++,-}$ and $\Delta^{+,0}$ obtained in the continuum
 limit are the same within statistical uncertainties.

The reliable determination of the lattice spacing from the nucleon mass
as well as the fact that isospin breaking is consistent with zero for
these lattices paves the way for further applications of twisted mass
fermions in the baryon sector.  

\section *{Acknowledgments}
We thank all other members of the ETM Collaboration for
very valuable discussions and for a most enjoyable and
fruitful collaboration.
We also thank Th. Hemmert for his comments on the nucleon $\sigma$-term.
G.K. is supported by the Cyprus Research Promotion Foundation under contract
$\Pi$ENEK/ENI$\Sigma$X/0505-39. 
This work is supported in part by  the DFG
Sonder\-for\-schungs\-be\-reich/Transregio SFB/TR9-03, DFG project
JA 674/5-1 and the EU Integrated
Infrastructure Initiative Hadron Physics (I3HP) under contract
RII3-CT-2004-506078.  
  R.F. acknowledges MIUR (Italy)
for partial financial support under the contracts PRIN04 and and PRIN06.

The analysis runs were performed in the computing centers of
the CRI (Paris-Sud), the CCIN2P3 (IN2P3), the CCRT (CEA). This work was 
partly done
with the help of the ``Project ANR-NT05-3-43577",
 which is a non thematic project
named QCDNEXT and funding received  from the
 Cyprus Research Promotion Foundation under contract EPYAN/0506/08. 
Computer time for this project is made available to us by the
John von Neumann Institute for Computing on the JUMP and Jubl systems
in J\"ulich and apeNEXT system in Zeuthen, 
by the UKQCD Collaboration
on the QCDOC machine at Edinburgh,
by INFN and CNRS on the apeNEXT systems in Rome,
by BSC on MareNostrum in Barcelona and by the Leibniz Computer
Center on the Altix system in Munich. 
We thank these Computer Centers and
their staff for technical advice and help.

\bibliography{NDmass_ref}

\end{document}